\documentclass[a4paper,10pt]{article}

\usepackage{epsfig}
\usepackage{psfig}

\usepackage{amssymb}

\usepackage{makeidx}  % allows for indexgeneration

\newtheorem{mydef}{\noindent {\sc Definition}} 
\newtheorem{myth}{\noindent {\sc Theorem}} 
 
\newtheorem{myprop}{\noindent {\sc Proposition}}

\def \s {\hspace{0.05cm}}
\def \reals {\mathbb{R}}

\def \aa {\mathbf{a}}
\def \bb {\mathbf{b}}
\def \xx {\mathbf{x}}
\def \yy {\mathbf{y}}
\def \uu {\mathbf{u}}

\def \zz {\mathbf{z}}
\def \ff {\mathbf{f}}

\def \projx {\pi_{_X}}

\def \acal {{\cal A}}
\def \ccal {{\cal C}}

\def \lcal {{\cal L}}

\def \scal {{\cal S}}

\def \vcal {{\cal V}}
\def \xcal {{\cal X}}
 
\def \foresigmatrans { {\s \stackrel{\sigma}  {\longrightarrow}\s } }
\def \backsigmatrans { {\s \stackrel{\sigma}  {\longleftarrow}\s } }

\title { Approximate Computation of Reach Sets in Hybrid Systems}

\author{ D. Ravi\thanks{Current Address: Faculty of Mathematics and Computer Science, The Weizmann Institute of Science,
    76100 Rehovot, Israel} \s and \s R. K. Shyamasundar\\ \\
  \small {School of Technology and Computer Science}\\
  \small{Tata Institute of Fundamental Research} \\
  \small{Homi Bhabha Road, Colaba, Mumbai 400 005, India}\\
  \small{ {\sf \{ravi, shyam\}@tcs.tifr.res.in} }  }

\date{}

\begin{document}
\maketitle

\begin{abstract}

%\noindent
 One of the most important problems in hybrid systems is the
  {\em reachability problem}. The reachability problem has been shown
  to be undecidable even for  a subclass of {\em linear}
  hybrid systems.  In view of this, the main focus in the area of hybrid systems
  has been to find {\em effective} semi-decision
  procedures for this problem. Such an algorithmic approach
  involves finding methods of computation and representation
  of reach sets of the continuous variables within a discrete
  state of a hybrid system. In this paper, after presenting
  a brief introduction to hybrid systems and reachability
  problem, we propose a computational method for obtaining
  the reach sets of continuous variables in a hybrid system.
  In addition to this, we also describe a new algorithm to
  over-approximate with polyhedra the reach sets of the
  continuous variables with linear dynamics and polyhedral
  initial set. We illustrate these algorithms with typical 
  interesting examples.

\end{abstract}

\section{Introduction}
 Hybrid systems combine discrete and continuous dynamics. The
 dynamics of the continuous variables within a discrete state
 are specified by differential equations or differential
 inclusions. An important problem in the analysis and sysnthesis
 of such systems is the so called {\em reachability problem},
 which asks, for two sets of configurations of a given hybrid system,
 say $\xcal_1$ and $\xcal_2$, where a configuration consists of
 discrete and continuous components, whether or not there is a
 hybrid trajectory with the initial configuration in $\xcal_1$
 and the final configuration in $\xcal_2$. A hybrid trajectory
 may be described as a trajectory of configurations consisting
 of discrete state jumps and smooth arcs, where each arc evolves
 according to the continuous dynamics of a discrete state, with
 the starting and end points of each arc satisfying the jump
 conditions of discrete transitions. A more precise definition is
 presented in Sec. 2, which provides a concise introduction to
 hybrid systems and the reachability problem.
\\

 The reachability problem is undecidable for certain classes
 of linear hybrid systems (i.e., hybrid systems having linear
 trajectories within each discrete state and linear or constant
 reset maps, also called {\em constant slope hybrid systems})
 \cite{Alur95, Kesten-JIC99}, although in some cases decidability
 results have been obtained \cite{Laff99a,Laff99b} (see also
 \cite{Kesten-JIC99}). Therefore, for a general hybrid system,
 a reasonable alternative appears to be to find semi-decision
 procedures for the reachability problem
 \cite{Henzinger-IEEE-AC98,Pappas-IEEE-CDC00,ABDM-IEEE00}.
 A computational approach to this problem also requires
 finding the set of states reached by the continuous variables
 evolving according to the dynamics of a discrete state. In this
 paper, we consider the problem of computing and representing the
 reach sets of the continuous variables within a discrete state
 when the dynamics of the continuous variables are specified by
 differential equations with initial conditions belonging to
 a specified initial set.
 \\

  In this context, various methods have been proposed in the literature
  for finding reach sets of continuous variables \cite{DM98,LTS,V98}.
  In Sec. 3, we describe a method for computing the reach sets based
  on the idea that a subset of the boundary of the initial set may
  be found, such that it is sufficient to compute the solutions with
  the initial points lying in that set. This method is similar to that
  in \cite{DM98} (and also to some extent to that described in \cite{LTS}).
  In particular, we present a schematic algorithm, which is somewhat
  more general in its scope than that in \cite{DM98} and simpler
  compared to that in \cite{LTS}.
 \\

% Formal analysis and verification of properties such as
% {\em safety, liveness, eventuality,} etc., often requires
% finding the set of phases reached by the continuous variables
% evolving according to the dynamics of a discrete state.

  Besides these algorithms for reach set computation, an
  equally important issue the representation of the reach
  sets for manipulating the sets efficiently. This requires
  representation of the reach sets in terms of more convenient
  sets such as, for example polyhedral or subalgebraic sets,
  that are simple to represent and easy to handle for practical
  purposes. However, since the representing class of sets may not
  contain a member that exactly equals the reach set, we may have
  to find approximations to the reach sets by those that belong
  to the representing class. (See, for example, \cite{V98,ABDM00}
  for approximation with polyhedra, and \cite{KV00} for approximation
  using ellipsoids). Typically, over-approximations may be used
  for verifying whether a safety requirement may potentially
  be violated by any of the behaviours starting from a given
  initial set, while under-approximations are needed for
  characterizing a set of states from which a desirable
  property is always achievable. We describe in Sec. 4 a
  method for over-approximating the reach sets by polyhedra
  when the dynamics are specified by linear differential
  equations and the initial set is a polyhedron. These
  algorithms are illustrated with some simple examples
  in Sec. 5, while Sec. 6 concludes the paper.

\section{A Brief Survey of Hybrid Systems}

  In this section, we present a brief introduction
 to hybrid systems, and provide motivation for the
 remaining sections.

\subsection{Preliminary Definitions}

We begin with a somewhat detailed but general definition of hybrid systems.
 
\begin{mydef}
\noindent  \label{HS} A {\em hybrid system}  is
 a tuple $H = (Q,X,\Sigma,G,E,Init,f)$, where
\begin{enumerate}
  \item $Q$ is a finite set of discrete {\em states}
             (also called {\em locations}).

  \item $X = \reals^n$, $n \geq 1$, is the set of continuous states,
        where $\reals$ denotes the set of real numbers. We demote the
        continuous state variable by $\xx$.

  \item $\Sigma$ is a finite set of discrete events or environment actions.
         Some events in $\Sigma$ are controllable, while others not. Hence,
         it is convenient to assume that $\Sigma = \Sigma_c \bigcup \Sigma_d$,
         where $\Sigma_c$ is the set of controllable events, and $\Sigma_d$ is
         the set of uncontrollable or disturbance events.
%
%         We denote
%         by $\perp$ the {\em empty} event, which may not be in $\Sigma$.
%         The occurrence of $\perp$ is only to allow the time to pass by
%         without any significant environment action that may affect the
%         behavior of the hybrid system. Further, let
%         $\Sigma^{\perp} = \Sigma \bigcup \{ \perp\}$ and 
%         $\Sigma_c^{\perp} = \Sigma_c \bigcup \{ \perp \} $.
%   

  \item $G \subseteq Q \times X$ is a set of state invariance conditions.
        When the system is in state $q$, the continuous variables belong
        to $G(q) = \{\xx\in X\, : \; (q,\xx) \in G\}$.

  \item $E \subseteq Q\times{\cal P}(\reals^n)\times \Sigma \times \{\reals^n\rightarrow\reals^n\}\times Q$ is the set of transition edges. An edge
$e \in E$, where $e = (q_e,X_e,\sigma_e,r_e,q'_e)$, is interpreted as follows:

  \begin{itemize}
      \item If the continuous state is in $X_e$ and the event $\sigma_e$
            occurs, then transition edge $e$ is enabled in state $q_e$.
            Thus $X_e$ is the set of switching points of the continuous
            variables from state $q_e$ to $q_e'$. The set $X_e$ is
            specified by a predicate, and is also called a {\em guard}.
            In case there are many transition edges simultaneously enabled, 
            then the system may select one of the edges nondeterministically.
 
      \item If a transition edge $e$ is selected by the system, then the
            continuous state is reset using the function $r_e$ when the
            system enters state $q_e'$. The reset values obey the state
            invariance condition, $r_e(X_e) \subseteq G(q_e')$. 

   \end{itemize}
%
%   If the reset is nondeterministic, then 
% $E \subseteq Q\times{\cal P}(\reals^n)\times \Sigma \times \{\reals^n\rightarrow 2^{\reals^n}\}\times Q$. In this case, if the edge $e$ is selected when the continuous
%      state is $\xx_e$, where $\xx_e \in X_e$, then, after entering state $q_e'$,
%      the continuous variables are nondeterministically reset to one of the values
%      in $r_e(\xx_e)$, and the reset values obey the state invariant condition
%      $r_e(\xx_e) \subseteq G(q_e')$, for each $\xx_e \in X_e$. However, unless
%      stated otherwise, we assume that the reset map is deterministic, and
%   $E \subseteq Q\times{\cal P}(\reals^n)\times \Sigma \times \{\reals^n\rightarrow\reals^n\}\times Q$.
%

  \item $Init \subseteq G$ is a set of initial conditions.

  \item $f$ is an $n$-dimensional vector field with real-valued components
  governing the dynamics of the continuous state $\xx$. The domain of definition
  of the function $f$ is a set ${\cal D}$, where ${\cal D} = Q \times X$ (if
  there are no continuous control variables) or ${\cal D} = Q \times X \times U$
  (if $U \subseteq \reals^m$, $m \geq 1$, is the range of the continuous
  control variable, which is a function $\uu : [0,T_q) \rightarrow U$, where
  $T_q > 0$ is as large as may be necessary, depending on the discrete state
  $q \in Q$). When in state $q$, the continuous variables evolve according
  to the dynamical law
 \[
      \frac{d\xx}{dt}  = f(q,\xx) 
 \]
  or according to the dynamical law
 \[
      \frac{d\xx}{dt}  = f(q,\xx, \uu) 
 \]
  depending on the presence or absence of the continuous control
  variables.\footnote{Another possibility is to specify the
  system dynamics of the continuous variables in terms of differential
  inclusions. In this case, $f(q,\xx)$ is a set, and the continuous
  variables evolve according to
 \[
      \frac{d\xx}{dt}  \in f(q,\xx).
 \]
  But we do not consider this case here.}

  The initial conditions of the continuous evolution are specified
  either by the reset maps when the system takes a discrete transition
  and reaches the state $q$, or by the (nondeterministic) initial
  conditions of the system given as the set $Init$.
  \end{enumerate}

  If, in addition to the set of initial values, $Init$, the set of final or
  accepting values in $Q\times X$, denoted by $Final$, are also specified,
  then the hybrid system is called a {\em hybrid automaton}.
\end{mydef}

  Referring to the system dynamics in the above definition, for our purposes,
  we assume that there are no continuous control variables. Therefore,
  in the sequel, we assume that the dynamics of the continuous variables
   are specified in the form
  \[
      \frac{d\xx}{dt}  = f(q,\xx)
  \]  
  without any continuous control variables.

\begin{mydef}
 \label{Config} Let $H = (Q,X,\Sigma,G,E,Init,f)$ be a hybrid system.
  Each point $(q,\xx) \in Q \times X$ is called a {\em configuration} of
  the hybrid system $H$.

\end{mydef}

  In Definitions \ref{Step} - \ref{PreSigma}, we introduce certain
  terminology, starting with the notion of a {\em step} of a hybrid
  system, leading upto the notions of {\em predecessors} and
  {\em successors} of a configuration $(q,\xx)$. But before that,
  we mention that predecessors and successors may be defined
  irrespective of initial conditions, whereas for defining
  the executions of a hybrid system, we require the specification
  of the initial set $Init$, as will be seen later (refer to
  Definitions \ref{Traject}, \ref{PostSigma} and \ref{PreSigma}) .

\begin{mydef}
\noindent  \label{Step}
 Let $H = (Q,X,\Sigma,G,E,Init,f)$ be a hybrid system, and let
 $(q,\xx)$ and $(q',\xx')$ be two configurations of $H$. Then,
 the pair of configurations $((q,\xx), (q',\xx'))$ is called
 \begin{enumerate}
  \item a {\em time-step}, if $q = q'$ and, for some $t \geq 0$,
   there is a function $\yy : [0, t] \rightarrow X$, satisfying
   $\dot{\yy}(s) = f(q,\yy(s))$, $\yy(s) \in G(q)$, for
   $s \in [0, t]$, $\yy(0) = \xx$ and $\yy(t) = \xx'$. 

  \item an {\em edge-step} if there is an edge
        $e = (q_e,X_e,\sigma_e,r_e,q_e') \in E$, such that
        $q = q_e$, $q' = q_e'$, $\xx \in X_e$ and $\xx' = r_e(\xx)$.
 
  \item a $\sigma$-{\em step}, where  $\sigma \in \Sigma$, if there is
        an edge $\ell = (q_\ell,X_\ell,\sigma_\ell,r_\ell,q_\ell') \in E$
        with $\sigma = \sigma_{\ell}$, such that
        $q = q_\ell$, $q' = q_\ell'$, $\xx \in X_\ell$ and $\xx' = r_\ell(\xx)$.        
 \end{enumerate}
  A {\em step} of the hybrid system $H$ is a pair of configurations
  $((q,\xx), (q',\xx'))$,  such that $((q,\xx), (q',\xx'))$ is either
  a time-step for some $t \geq 0$, or an edge-step for some $e \in E$,
  or a $\sigma$-step for some $\sigma \in \Sigma$.

\end{mydef}

  Of course, every edge-step is a $\sigma$-step for some
  $\sigma \in \Sigma$, and conversely, every $\sigma$-step
  is an edge-step for some edge $e \in E$. Hence these
  definitions may seem redundant. However, the distinction
  between the two types of transitions would become obvious
  if there is a {\em discrete state controller}, defined as
  a function $C : Q\times X \rightarrow \Sigma_c$, that
  triggers discrete controllable events to enable a particular
  transition depending on the present configuration
  (see for instance \cite{STACS95} for the case of 
  {\em timed automata}). But in this work, we shall not
  have occasion to discuss about discrete state controllers.

\begin{mydef} {\bf (Trajectories and Executions of a Hybrid System)}
\label{Traject} A {\em hybrid trajectory} or simply a {\em trajectory}
 of $H$ is a sequence of configurations
 $(q_1,\xx_1), (q_2,\xx_2), (q_3,\xx_3), \ldots$, where for
  each $i \geq 1$, $((q_i,\xx_i),(q_{i+1},\xx_{i+1}))$ is a step.
  A trajectory is
  \begin{enumerate}
   \item  finite, if the number of steps is finite; 

   \item  an {\em execution}, if $(q_1,\xx_1) \in Init$,
          where $(q_1, \xx_1)$ is the initial configuration; and

  \item  a {\em finite execution}, if it is both.
  \end{enumerate}

\end{mydef}

 We now define, for each $\sigma \in \Sigma$, the set-valued successor
function, $Post_\sigma : Q \times X \rightarrow 2^Q \times 2^X$.  As will
be briefly mentioned later (after Definition \ref{PreSigma}), our definitions
lead in a natural way to extract a {\em labeled transition system} (see
\cite{Henzinger-ICALP95}) from a hybrid system.

\begin{mydef}
\noindent \label{PostSigma}   {\bf ($\sigma$-Successors of a Configuration)}\\
 Let $H = (Q,X,\Sigma,G,E,Init,f)$ be a hybrid system, $\sigma \in \Sigma$ and
 $(q,\xx) \in Q \times X$. Then, we define $Post_\sigma (q,\xx)$ to be the
 union of the two sets $S_1, S_2^\sigma \subseteq Q\times X$, where
 
  \begin{enumerate}
    \item
 $S_1 = \{ (q',\xx') : ((q,\xx),(q',\xx')) \textrm{ is a time-step for some $t \geq 0$} \}$; and
    \item
 $S_2^\sigma = \{ (q'',\xx'') : ((q',\xx'),(q'',\xx'')) \textrm{ is a $\sigma$-step  for some  $(q',\xx') \in S_1$ } \}$.
  \end{enumerate}
  Further, if $(q_s,\xx_s) \in Post_{\sigma}(q,\xx)$, then $(q_s,\xx_s)$
 is called a {\em $\sigma$-successor} of $(q,\xx)$.
 
\end{mydef}

  The set $S_1$, as in the first part of this definition, is of main
  interest to us in the later sections of this paper. Specifically,
  we will deal with the set $Reach_q^G\left(X_0\right)$, where
  $q \in Q$ and $X_0 \subseteq G(q)$, defined as
\begin{equation}
 Reach_q^G\left(X_0\right) =
     \left \{ \xx : ((q,\xx_0), (q,\xx)) \textrm{ is a time-step for some}
    \textrm{ $t \geq 0$ and $\xx_0 \in X_0$ } \right \}.
   \label{Reach_q}
\end{equation}
 In the subsequent sections, we will be concerned more with this and
 a related set. But for now, we shall proceed with our discussion
 with the following definition:
%
% In the subsequent sections, we shall be mainly concerned with the
% computational aspects,  and related issues such as approximations,
% of the set $ Reach_q^G\left(X_0\right)$ and a small varient of this
% set (without the global invariance condition).
%                     

\begin{mydef}
\noindent \label{PreSigma}   {\bf ($\sigma$-Predecessors of a Configuration)}\\
 Let $H = (Q,X,\Sigma,G,E,Init,f)$ be a hybrid system, $\sigma \in \Sigma$ and
 $(q,\xx) \in Q \times X$. Then, we define $Pre_\sigma (q,\xx)$ to be the
 union of the two sets $P_1, P_2^\sigma \subseteq Q\times X$, where
 
  \begin{enumerate}
    \item
 $P_1 = \{ (q',\xx') : ((q',\xx'),(q,\xx)) \textrm{ is a time-step for some $t \ge0$} \}$; and
    \item
 $P_2^\sigma = \{ (q'',\xx'') : ((q'',\xx''),(q',\xx')) \textrm{ is a $\sigma$-step  for some  $(q',\xx') \in P_1$ } \}$.
  \end{enumerate}
  Further, if $(q_p,\xx_p) \in Pre_{\sigma}(q,\xx)$, then $(q_p,\xx_p)$
 is called a {\em $\sigma$-predecessor} of $(q,\xx)$.
 
\end{mydef}                                                                       

 Note that the notion of $\sigma$-successor generalizes the notion
 of $\sigma$-step by including time-steps. The set valued function
  $Post_{\sigma}(q,\xx)$ defines, in a natural way,
 a transition relation, 
 $\foresigmatrans \subseteq (Q\times X) \times (Q\times X)$,
 as follows: $(q,\xx) \foresigmatrans (q',\xx')$,
 if $(q',\xx') \in Post_{\sigma}(q,\xx)$.
 The transition relation $\foresigmatrans$ is also
 called {\em $\sigma$-transition} relation.
 Similarly, the set valued function $Pre_{\sigma}(q,\xx)$ defines
 another transition relation,  
 $\backsigmatrans \subseteq (Q\times X) \times (Q\times X)$, as follows:
 $(q,\xx) \backsigmatrans (q',\xx')$, if
 $(q,\xx) \in Pre_{\sigma}(q',\xx')$.
\\

 For two configurations  $(q_1, \xx_1)$ and $(q_2, \xx_2)$,
 if $((q_1, \xx_1), (q_2, \xx_2))$ is a $\sigma$-step, then
 $(q_2, \xx_2) \in Post(q_1, \xx_1)$ and $(q_1, \xx_1) \in Pre(q_2, \xx_2)$;
 hence, in this case, we have $(q_1, \xx_1) \foresigmatrans (q_2, \xx_2)$ and
 $(q_1, \xx_1) \backsigmatrans (q_2,\xx_2)$. This may mislead us to get
 the false impression that $\foresigmatrans$ and $\backsigmatrans$
 are the same; and to avoid any such possible confusion, we emphasize
 that, in general, it is not true that
 $(q_1, \xx_1) \foresigmatrans (q_2,\xx_2)$ implies
 $(q_1, \xx_1) \backsigmatrans (q_2,\xx_2)$; and the same
 with the converse statement. Hence,
 the two relations $\foresigmatrans$ and $\backsigmatrans$ are
 not the same.
\\

The definitions of $Post_\sigma$ and $Pre_\sigma$ can be extended
straightforwardly to subsets of $Q\times X$, as follows:
for $S \subseteq Q \times X$, define

\[
  Post_\sigma(S) = \bigcup_{(q,\xx) \in S} Post_{\sigma}(q,\xx), ~ \textrm{ and }
 ~ Pre_\sigma(S) = \bigcup_{(q,\xx) \in S} Pre_{\sigma}(q,\xx).
\]
  Finally, define
\[
  Post(S) = \bigcup_{\sigma\in \Sigma} Post_{\sigma}(S), ~ \textrm{ and } ~
  Pre(S) = \bigcup_{\sigma\in \Sigma} Pre_{\sigma}(S).
\]
We sometimes refer to $Post$ and $Pre$ as the {\em 1-step} transition
functions\footnote{This should not be confused with the notion of
{\em step} as defined in Definition \ref{Step}.}. More generally, the 
{\em $k$-step} transition functions $Post^k$ and $Pre^k$ are defined
 inductively as follows: For $S \subseteq Q\times X$,
\[
 Post^1(S) = Post(S), ~\textrm{ and  for $k \geq 2$, } ~ Post^k(S) = Post(Post^{k-1}(S)).
\]
Similarly,
\[
 Pre^1(S) = Pre(S), ~\textrm{ and for $k \geq 2$, } ~ Pre^k(S) = Pre(Pre^{k-1}(S)).
\]  
Finally, define $Post^*$ and $Pre^*$ as
\[
  Post^*(S) = \bigcup_{k \geq 1 } Post^k(S), ~ \textrm{ and } ~
  Pre^*(S) = \bigcup_{k \geq 1 } Pre^k(S).
\]     

\subsection{ \label{ReachProblem} Reachability Problem for Hybrid Systems}

\noindent In the notation just discussed, the
 {\em reachability problem} for a hybrid system
$H = (Q,X,\Sigma,G,E,Init,f)$, where $Init$ is
not necessarily specified in advance, may be posed
as follows:
\\
 
  \begin{tabular}{|l|}
  \hline
  { \bf ReachProblem1:} \s
 \sf{For two subsets $S_1$ and $S_2$ of $Q \times X$, is there a finite}\\
 \hspace{1.2in}  \sf{ trajectory, $(q_1,\xx_1), (q_2,\xx_2), \ldots, (q_N, \xx_N)$, for some }\\
\hspace{1.2in} \sf { $N \geq 1$, such that $(q_1,\xx_1) \in S_1$ and $(q_N,\xx_N) \in S_2$ ?} \\
\hline
\end{tabular}
\\

\noindent This may also be rephrased as 
\\
 
  \begin{tabular}{|ll|}
  \hline
 { \bf ReachProblem2:} &
 \sf{For two subsets $S_1$ and $S_2$ of  $Q \times X$}, \\
 & \sf { whether $Post^*(S_1) \bigcap S_2 \neq \emptyset$ ?} \\
 & \em {equivalently:}  \sf { whether $S_1 \bigcap Pre^*(S_2) \neq \emptyset$ ?} \\
\hline
\end{tabular}
\\

\noindent If the initial set $Init$ is specified in advance, then, with
$S_1 = Init$, the reachability problem {\bf ReachProblem1} becomes
\\

  \begin{tabular}{|ll|}
  \hline
 {\bf ReachProblem3:} &  \sf{For a subset $S$ of  $Q \times X$,}\\
                      &  \sf{is there a finite execution of $H$}\\
                      & \sf{with final configuration in $S$ ?}\\
  \hline                                                                          \end{tabular}
\\
\\

  There is a counterpart to the reachability problem, called the
 {\em avoidance problem}, which may be posed as follows:
\\
 
  \begin{tabular}{|ll|}
  \hline
{ \bf AvoidProblem:} &
 \sf{For two subsets $S_1$ and $S_2$ of  $Q \times X$}, \\
 & \sf { whether $Post^*(S_1) \bigcap S_2 = \emptyset$ ?} \\
 & \em {equivalently:}  \sf { whether $S_1 \bigcap Pre^*(S_2) = \emptyset$ ?} \\
\hline
\end{tabular}
\\                                                                                
\\

 In the sequel, we restrict our attention to {\bf ReachProblem1} or
{\bf ReachProblem2}. We observe that the answers to these
questions depend, in general, not only on the hybrid
system $H$, but also on the class ${\cal C}$ of subsets
of $Q\times X$ that are under consideration. Informally,
the class ${\cal C}$ is required to 
\begin{enumerate}
  \item include a specified class of sets ${\cal S}$, where ${\cal S}$ 
        may consist of the initial set, sets defined by state invariance
        conditions, those defined by guard conditions, and domains and
        ranges of reset maps of edges (and also final sets, if specified),

  \item be closed under the boolean set operations of union and
        complimentation, and under the functions $Post_\sigma$ and
        $Pre_\sigma$, and

  \item have an effective decision procedure for answering
        questions such as, for two sets $S_1, S_2 \in {\cal C}$,
        whether $S_1 = S_2$ or not.

\end{enumerate}

  A formal presentation of these notions is beyond the scope of
  this work, although a brief discussion may be found in Appendix A.
  For more details, the interested reader may refer to the references
  on Model Theory, such as \cite{Hodges, Marker, vanDalen, Tarski51}.
  But, for our purposes, we will be content with the following
  (somewhat informal) definition:
  
\begin{mydef}
  An algorithm for the reachability problem is said to be
  \begin{enumerate}
    \item a {\em decision procedure}, if the algorithm stops
     after a finite number of steps with the correct answer,
     where the answer can be either {\em yes} or {\em no}.

    \item a {\em semi-decision procedure}, if the algorithm
     \begin{enumerate}
       \item never stops with an incorrect answer, and
       \item always stops after a finite number of steps with the
             correct answer, whenever the answer is {\em yes}.
      \end{enumerate}
  \end{enumerate}
 \noindent A hybrid system $H$ is {\em decidable}, if there is a decision
          procedure for the reachability problem for $H$.
\end{mydef}

 Obviously, for a hybrid system $H$, if there are semi-decision
 procedures for both reachability problem and avoidance problem,
 then $H$ is decidable.
\\

  We describe below a schematic semi-decision procedure for the
 reachability problem of a hybrid system:
\\

\begin{tabular}{|l|}
  \hline
   {\sf Semi-Decision Procedure for} {\bf ReachProblem2 } \\
\\

   \begin{tabular}{lcl}
     \indent   {\bf Input} & {\bf :} &  {\sf Sets $S_1$ and $S_2$} \\
     \indent   {\bf Output}  & {\bf :}  & {\sf ``yes'', if there is a trajectory from $S_1$ to $S_2$} \\
   \end{tabular}
\\
\\

 \indent \indent   {\bf Initialization} $S := S_1$ \\
\\
 
 \indent \indent   {\bf while} $S \bigcap S_2 = \emptyset$ {\bf do} \\

 \indent  \indent  \indent $S := Post(S)$ \\
 
\indent  \indent {\bf end while} \\
\\

\indent  \indent {\bf return} {\sf ``yes''}\\
 
  \hline                                                                           

\end{tabular}
\\
         
 In the later part of this section, we shall be mainly concerned with the
 computational aspects of the $Post$-operator appearing in the while-loop
 of the above schematic.

\subsection{Computation of the $Post$-operator}

  We describe here a computational approach to finding $Post(S)$,
  $S \subseteq Q\times\reals^n$, as in the semi-decision procedure
  for {\bf ReachProblem2} discussed in Sec. \ref{ReachProblem}.
  Recall that, within a discrete state $q \in Q$, the continuous
  dynamics of $\xx$ are specified by
\begin{equation}
     \frac{d\xx}{dt}   =  f(q, \xx) \label{Eqn1}
\end{equation}
with the initial conditions $\xx(0) = \xx_0 \in X_0 \subseteq X = \reals^n$.
We begin our discussion with the following definition.

\begin{mydef} \label{FlowPhi}
 Let $\reals^+$ denote the set of nonnegative real numbers. Then,
 for each discrete state $q \in Q$, the {\em flow} associated with
 equation (\ref{Eqn1}) is a  function
 $\phi_q:\reals^n \times \reals^+ \rightarrow \reals^n$,
 defined as $\phi_q(\xx,t) = \gamma_\xx(t)$, $t \geq 0$ and
 $\xx \in \reals^n$, where the function
 $\gamma_\xx : \reals^+ \rightarrow \reals^n$ satisfies
\[
\frac{d\gamma_{\xx}(t)}{dt} =  f(q, \gamma_\xx(t) ),
\]
 with the initial condition $\gamma_\xx(0) = \xx$.
\end{mydef}

 Recall the set of reachable phases $Reach_q^G$ as defined in 
(\ref{Reach_q}). In the following definition, the same is defined
in terms of the flow function $\phi_q$:

\begin{mydef} 
 Let $q \in Q$ and $X_0 \subset \reals^n$. Then, the set of
 reachable continuous phases in the state $q$, is the set
 $Reach_q^G\left(X_0\right)$ defined as 
\begin{eqnarray}
  Reach_q^G(X_0) & = & \left \{ \xx \in \reals^n : ~ \exists \xx_0 \in X_0\;
  \exists t \geq 0 \textrm{  $\xx = \phi_q(\xx_0, t)$ and} \right.  \nonumber \\
  &&      \indent  \left. \forall s
   ~ 0 \leq s \leq t \Rightarrow \phi_q(\xx_0, s) \in G(q) \right \}
   \label{Reach_G_q}
\end{eqnarray}
\end{mydef}

\noindent We now define another operator, called the {\em projection} operator,
as follows:
\begin{mydef} \label{ProjX}
Let $S \subseteq Q \times X$. Then, the {\em projection of S on X},
is the set $\projx (S) = \{\xx \in X :\; (q,\xx) \in S\}$.
\end{mydef}

\noindent In terms of the operators $\projx $ and $Reach_q^G$, the operator
$Post$ may be rewritten as
\begin{equation}
Post (S) =
  \mathop{\bigcup_{ e = (q_e,X_e,\sigma_e,r_e,q'_e) \in  E, \; q = q_e} }_{X_e \bigcap \; [ Reach_q^G \; \circ \; \projx  (S) ]\s \neq \emptyset}  ~~
  \{q_e'\} \times  [ r_e \s \circ \s Reach_q^G \s \circ \s \projx   (S)],
      \label{PostOperator}
\end{equation}
where, for two operators $g$ and $h$, $g\circ h$ is the composition of
the two operators. With regard to computational issues, the set operators
$\projx$ and $r_e$ may not pose difficulties (provided $r_e$ is suitably
specified). Hence the problem reduces to that of computation of $Reach_q^G$,
and effectively representing the resulting set. Referring to the definition
of $Reach_q^G$ as in (\ref{Reach_G_q}), the significant and challenging task
is the elimination of quantifiers, whenever possible. But, as shown in
\cite{vanDenDries}, not all theories of the real number system may admit
quantifier elimination (see also the discussion presented in
\cite{PappasPhDThesis} and in Appendix A). This fact provides a motivation
for the study of alternative methods (without requiring quantifier
elimination method) for computating the set $Reach_q^G(X_0)$.
\\

Specifically, in the remaining part of the work, we shall be concerned
with computation of the operator $Reach_q^G$ and with representation of
the resulting sets. Below is a concise description of the questions that
we shall be interested in and of the organization of the rest of the paper:

 \begin{enumerate}

  \item How to effectively compute the set operator $Reach_q^G$, at least
       when the global invariance condition is not imposed but a time bound
       is specified? In this case, we are interested in the set 
        $Reach_q(X_0,[0,\tau])$,  $\tau \geq 0$, $X_0 \subset \reals^n$,
       defined as
       \[
            Reach_q (X_0, [0,\tau]) = \left \{ \xx \in \reals^n :
               ~ \exists \xx_0 \in X_0\; \exists t \;  0 \leq t \leq \tau
               \textrm { and }  \xx = \phi_q(\xx_0, t)   \right \}.
       \]
        In Sec. 3, we present a schematic algorithm for computing
        $ Reach_q (X_0, [0,\tau])$. This algorithm is based on a
         generalization of the method described in \cite{DM98}.

  \item How to compute -- either exactly or approximately -- the set
        operator $Reach_q^G$ when the global invariance condition is
        imposed (without time bound). In this case, we have to deal 
        with the set operator $Reach_q^G$ as defined in (\ref{Reach_G_q}),
        and obviously, it would be best to find an algorithm based on
        quantifier elimination. But, as mentioned earlier, we do not
        assume that quantifier elimination method is feasible, hence we
        may have to find an algorithm that computes an approximation to
        $Reach_q^G$. In Sec. 3, we indicate how to extend the algorithm
        for computing $Reach_q(X_0,[0,\tau])$, as mentioned above,
        to compute an under-approximation or an over-approximation
        (depending on which one is preferred) to the set $Reach_q^G(X_0)$.

  \item Finally, how to represent the reach sets obtained by the operators
        $Reach_q$ and $Reach_q^G$, i.e. the sets $Reach_q(X_0,[0,\tau])$
        or $Reach_q^G(X_0)$ as the case may be, such that boolean set operations
        such as union and complementation can be performed efficiently. 
        This requires represention of the reach sets in terms of sets that
        are simple and easy to handle, such as polyhedral sets and subalgebraic
        sets. In \cite{ABDM00}, an algorithm algorithm for over-approximating the
        reach sets using polyhedral sets is presented. In Sec. 4, we describe
        another algorithm for over-approximating the reach sets with polyhedral
        sets. The  algorithm presented in this paper differs from that of
        \cite{ABDM00}, but if these two methods -- that described in Sec. 4
        and that of \cite{ABDM00} -- are used together, then better results
        of over-approximation may be obtained.

\end{enumerate}

  The remaining part of the paper is orgainized as follows. In Sec. 3 and
  Sec. 4, we restrict our attention to these issues. We illustrate these
  algorithms with simple examples in Sec. 5, while Sec. 6 summarizes the paper.

\section{Computational Approaches for Finding Reach Sets}

\subsection{Preliminary Discussion on Reach Sets}

Let $H = (Q,X,\Sigma,G,E,Init,f)$ be a hybrid system. For
$q \in Q$, consider the equation
\begin{equation}
     \frac{d\xx}{dt}   =  f(q, \xx) \label{Eqn1}
\end{equation}
with the initial conditions $\xx(0) = \xx_0 \in X_0 \subset X = \reals^n$.
It is customary to assume that $f(q, \cdot)$ is Lipschitz continuous in
the second variable, in order to ensure existence and uniquness of
solutions to the system of differential equations in (\ref{Eqn1}). Further,
we assume that the initial set $X_0$ is closed. Let, as before, the flow
(refer to Defintion \ref{FlowPhi}) associated with (\ref{Eqn1}) be $\phi_q(\xx,t)$
and let $Reach_q(X_0,t) = \{\phi_q(\xx_0,t) : \xx_0 \in X_0\}$
be the set of phases reached at time $t \geq 0$ in state
$q$ with initial conditions in $X_0$. Further, let
\begin{eqnarray*}
Reach_q(X_0,[0,\tau]) & = & \bigcup_{0\leq t \leq \tau} Reach_q(X_0,t),~~~~ \tau \geq 0,\\
Reach_q(X_0,[0,\infty)) & = &  \bigcup_{\tau > 0} Reach_q(X_0,[0,\tau]).
\end{eqnarray*}
We shall now define, for a set $X_q$, such that 
$X_0 \subseteq X_q \subseteq \reals^n$, another set
$Reach_q'\left(X_0, X_q, \, [0,\infty)\right)$.
For this purpose, we first define, for each $\xx \in X_0$,
a number $\tau_{\xx}$ as follows:
\begin{equation}
 \tau_{\xx} = \inf\{t > 0 \s: \phi_q(\xx,t) \not \in X_q\},  \label{taux_inf}
\end{equation}
where we assume that if $\tau_{\xx} = \infty$, then
$\phi_q(\xx,\infty)$ denotes the set of all $\omega$-limit points
(see, for instance, \cite{Hirsch&Smale}) of $\phi_q(\xx,t)$,
for $t \geq 0$, and
 \[
   \{ \phi_q(\xx,t): 0 \leq t \leq \infty\}  =
   \{\phi_q(\xx,t): 0 \leq t < \infty\} \bigcup \phi(\xx,\infty).
  \]         
Further, let $\Theta_q(\xx, X_q) \subseteq \reals^n$ be defined as follows:
\[
   \Theta_q(\xx, X_q) = 
  \left \{
   \begin{array}{l}
       \{ \phi_q(\xx,t)\s: 0 \leq t \leq \tau_{\xx} \}, ~~\textrm{if $\tau_{\xx} < \infty$ and $\phi(\xx,\tau_{\xx}) \in X_q$,} \\
       \{ \phi_q(\xx,t)\s: 0 \leq t \leq \infty \}, ~~\textrm{if $\tau_{\xx} = \infty$ and $\phi(\xx,\infty) \subseteq X_q$,} \\
       \{ \phi_q(\xx,t)\s: 0 \leq t < \tau_{\xx} \}, ~~\textrm{otherwise.}
   \end{array}
  \right.
\]
Then the set $Reach_q'\left(X_0, X_q, \, [0,\infty)\right)$ is defined as
\[
   Reach_q'\left(X_0, X_q, \, [0,\infty)\right) =
    \bigcup_{\xx \in X_0} \Theta_q(\xx, X_q).
\]
Now, by taking $X_q = G(q)$ in the above definition, we have
\[
  Reach_q^G\left(X_0\right) = Reach_q'\left(X_0, X_q, \, [0,\infty)\right), 
\]
where $Reach_q^G\left(X_0\right)$ is defined as in (\ref{Reach_G_q}), 
reproduced below for convenience:
\begin{eqnarray*}
  Reach_q^G(X_0) & = & \left \{ \xx \in \reals^n : ~ \exists \xx_0 \in X_0\;
  \exists t \geq 0 \textrm{  $\xx = \phi_q(\xx_0, t)$ and} \right.   \\
  &&      \indent  \left. \forall s
   ~ 0 \leq s \leq t \Rightarrow \phi_q(\xx_0, s) \in G(q) \right \}
\end{eqnarray*}

Throughout the rest of the discussion, we fix a state $q$, and
consider the problem of computing the sets $Reach_q(X_0,[0,\tau])$
and $Reach_q'\left(X_0, X_q, \, [0,\infty)\right)$.

% Since the discrete state $q$ is fixed, we denote, for convenience,
% $f(q, \xx)$ by $\ff(\xx)$, $\phi_q(\xx, t)$ by $\phi(\xx,t)$,etc.

\subsection{A Computational Method: Generalized Face-Lifting Algorithm}

When $X_0$ is suitably specified (such as, for example, a rectangle), 
this problem may be solved by finding the solutions to
(\ref{Eqn1}) with $\xx(0)$ on the boundary of $X_0$. This
results in evolving the boundary of $X_0$. More precisely,
let $S_0$ be the boundary of $X_0$, and 
$X(\tau) = X_0 \bigcup_{0\leq t \leq \tau} \{\phi_q(\xx_0,t) : \xx_0\in S_0\}$.
Then $X(\tau) = Reach_q(X_0, [0,\tau])$ (see Appendix A for a proof).
This is the idea underlying the computational approach, called
{\em face lifting} method, described in \cite{DM98}. In this section,
we shall study this in considerably more general setting, and describe
an algorithm, called {\em generalized face lifting} method. It may
be noted that the method described in \cite{DM98} does not assume
that a global invarience requirement is imposed; hence the algorithm
of \cite{DM98} computes only $Reach_q(X_0, [0,\tau])$. However, we
shall extend our method for computing $Reach_q(X_0, [0,\tau])$ for 
computing an approximation to the set $Reach_q^G(X_0)$, where the
approximation can chosen to be either under-approximation or
over-approximation.
\\

  Further, let $S_0^+ \subseteq S_0$ be the set of those boundary points
of $X_0$, such that the solution with initial point in $S_0^+$ extends into
 $X_0^c \; {\small {\stackrel{\Delta}{=}} }\; \reals^n \backslash X_0$ ,
 i.e.,
\begin{equation}
  S_0^+ = \{ \xx_0 \in S_0\s: \exists \epsilon \; =
      \; \epsilon(\xx_0) > 0 \textrm{ such that }
    \phi_q(\xx_0,t) \in X_0^c, ~ \forall t \in (0,\epsilon) \},
        \label{S_0_plus}
\end{equation}
and define
$X^+(\tau) = X_0 \bigcup_{0\leq t \leq \tau} \{\phi_q(\xx_0,t)\s :\; \xx_0\in S_0^+\}$.
It turns out that $X^+(\tau) = X(\tau) = Reach_q(X_0,[0,\tau])$ (see Appendix A).
Therefore, in order to find $Reach_q(X_0,[0,\tau])$, we have to
find only those solutions for which the initial conditions are
in $S_0^+$.
\\

  If $S_0^+$ can be found explicitly (by inspection of $\ff$ and $S_0$),
  then the problem reduces justifiably to finding the solutions with
  initial conditions in $S_0^+$. Otherwise, we may have to find a means
  of obtaining an outer approximation to $S_0^+$, i.e., a set $S_1^+$
  satisfying $S_0^+ \subseteq S_1^+ \subseteq S_0$. We suggest a way to find
  such a set. For this purpose, we assume that $X_0$ is specified as
  follows: there is a continuously differentiable function,
  $\ell : \reals^n \rightarrow \reals$, such that if
  $\xx \in {\stackrel{\circ}{X_0}}$ then $\ell(\xx) < 0$
  (where ${\stackrel{\circ}{X_0}}$ denotes the interior of $X_0$,
  i.e., the largest open set contained in $X_0$), and if
  $\xx \in X_0^c$ then $\ell(\xx) > 0$. Hence $\ell(\xx) = 0$ on $S_0$.
  Define
  \[
      S_1^+ = \{\xx \in S_0\s: ~ \nabla \ell (\xx) \cdot f(q,\xx) \geq 0 \}.
  \]
  It may be easily shown that $S_0^+ \subseteq S_1^+$ (see Appendix A).
  Futher, we observe that the set of points for which
  $\nabla \ell (\xx) \cdot f(q,\xx) > 0$, constitutes an
  inner approximation to $S_0^+$. The method of finding an $S_1^+$
  as described above can be extended easily to the situation where
  $X_0$ is specified as the set of intersection of finitely many
  sets of the form $\ell_i(\xx) \leq 0$, with the strict 
  inequality in ${\stackrel{\circ}{X_0}}$.
\\

  With this notation, we proceed to describe a schematic
  algorithm for finding $X^+(\tau)$ for $\tau \geq 0$. In
  order to exploit the semigroup property of the reach set, i.e., 
  $Reach_q(X_0,[0,t]) = Reach_q(Reach_q(X_0,[0,s]),[0,t-s])$, for any
  $t$ and $s$, with $0 \leq s \leq t$, we consider
  a sequence of time intervals $[0, \tau_1]$, $[\tau_1,\tau_2]$,
  $[\tau_2,\tau_3]$, etc., where
  $0 < \tau_1 < \tau_2 < \tau_3 < \ldots$, and
  $\tau_i \rightarrow \infty$, as $i \rightarrow \infty$.
  It may be convenient to choose for some $\tau > 0$,
  $\tau_i = i \tau$, $i = 0, 1, 2, \ldots$, although
  we do not require such an assumption in the algorithm.
\\

  \begin{tabular}{|l|}
  \hline
\\
   {\sf  Procedure for $Reach_q\left(X_0,[0,\tau]\right)$}\\
\\
    
   ~~{\bf initialize:}  $X^+(0) := X_0$, $F_0 := S_1^+$ and $i := 0$~~ \\

    ~~{\bf while} $\tau > \tau_i$ {\bf and} $F_i \neq \emptyset$ {\bf do} ~~\\

     \indent {\bf if} $\tau < \tau_{i+1}$ {\bf then} $\Delta_i := \tau-\tau_i$
                  {\bf else}  $\Delta_i := \tau_{i+1}-\tau_i$~ {\bf end if}\\

     \indent $T_i := \{\phi_q(\xx,t)\,:~t \in [0,\Delta_i], ~ \xx \in F_i \}$~~\\
   \indent \indent /* this computational step requires special attention! */ \\

     \indent $X^+(\tau_i+\Delta_i) := X^+(\tau_i) \bigcup T_i$\\

     \indent $F_{i+1} := \{\phi(\xx,\Delta_i)\,:~ \xx \in F_i \} \backslash X^+(\tau_i)$~~\\
    
     \indent $i := i+1$~~\\

    ~~{\bf end while}~~\\

\\
\hline
  \end{tabular}
\\
\\

  In the above schematic, the initialization step ``$F_0 := S_1^+$''
  could be replaced with ``$F_0 := S_0^+$'', as it is not necessary
  to initialize $F_0$ to $S_1^+$.  However, before proceeding further,
  it must be mentioned that with reference to this algorithm, we
  assume that the computational steps
  ``$T_i := \{\phi_q(\xx,t)\,:~t \in [0,\Delta_i], ~ \xx \in F_i \}$''
  and
  ``$F_{i+1} := \{\phi(\xx,\Delta_i)\,:~ \xx \in F_i \} \backslash X^+(\tau_i)$''
  can be performed effectively.
\\

  We now describe a method for extending this procedure to
  another procedure that computes an approximation
  to $Reach_q'(X_0,X_q,[0,\infty))$.  But since 
  $Reach_q^G(X_0) = Reach'(X_0,G(q),[0,\infty))$, the procedure
  to be described below finds an approximation to $Reach_q^G(X_0)$,
  when $X_q = G(q)$.  The approximation can be chosen to be either
  under-approximation or over-approximation, depeding on a flag
  ``$under\_approximate$'', passed as input to the algorithm
  (refer to the schematic algorithm described below). 
\\

 To facilitate the discussion, we consider
  the flow $\psi_q(\xx, t)$, $t \geq 0$, $\xx \in X = \reals^n$,
  corresponding to the differential equation
  \begin{equation}
     \frac{d\xx}{dt}   =  -f(q, \xx) \label{Eqn2}
  \end{equation}
  with the initial conditions $\xx(0) = \xx_0 \in  X = \reals^n$.
  Now since the flow $\psi_q$ has the opposite direction to that
  of $\phi_q$, for any $t \geq 0$ and two subsets $X_1$ and
  $X_2$ of $ \reals^n$, we have, $X_2 = \phi_q(X_1,t)$ if and
  only if $X_1 = \psi_q(X_2,t)$. The function $\psi_q$ will be
  used in the procedure for computing an approximation to
  $Reach'(X_0,X_q,[0,\infty))$, for finding, in each iteration,
  the set of initial conditions in $F_i$ corresponding to
  which the solutions for the time interval $[0,\Delta_i]$
  may violate the global invariance requirement. With this,
  we present the schematic algorithm as follows:
\\

  \begin{tabular}{|l|}
  \hline
\\
   {\sf  Procedure for $Reach_q'\left(X_0,X_q,[0,\infty)\right)$}\\
\\

    {\bf precondition: $ X_0 \subseteq X_q$}\\
    {\bf boolean input flag:  $under\_approximate$}\\
    \indent  /* $under\_approximate = 1$, if under-approximation is preferred */\\
    \indent  /*  by default, the procedure over-approximates the reach set */\\
\\

   ~~{\bf initialize:}  $X^+(0) := X_0$, $F_0 := S_1^+$ and $i := 0$~~ \\
 
    ~~{\bf while} $F_i \neq \emptyset$ {\bf do} ~~\\
 
      \indent $\Delta_i := \tau_{i+1}-\tau_i$~~\\
 
     \indent $T_i := \{\phi_q(\xx,t)\,:~t \in [0,\Delta_i], ~ \xx \in F_i \}$~~\\
\\

      \indent {\bf if} $T_i \subseteq X_q$ {\bf then} ~~ /* global invariance not violated */\\
       \indent \indent  $X^+(\tau_i+\Delta_i) := X^+(\tau_i) \bigcup T_i$\\
 
       \indent \indent  $F_{i+1} := \{\phi_q(\xx,\Delta_i)\,:~ \xx \in F_i \} \backslash X^+(\tau_i)$~~\\
 
      \indent {\bf else}      ~~~  /* global invariance violated by at least one trajectory */\\

       \indent \indent  $U_i := T_i \backslash X_q $ \\

      \indent \indent  $V_i := \{ \psi_q(\xx,t) :~ t \in [0,\Delta_i], ~ \xx \in U_i \}$ \\

       \indent \indent $F_i' = F_i \backslash V_i$ \\

      \indent \indent  $T_i' := \{\phi_q(\xx,t)\,:~t \in [0,\Delta_i], ~ \xx \in F_i'  \}$~~\\
\\

       \indent \indent   {\bf if} $under\_approximate = 1$ {\bf then} \\

       \indent \indent \indent  $X^+(\tau_i+\Delta_i) := X^+(\tau_i) \bigcup U_i$\\
 
       \indent \indent {\bf else} \\
             \indent \indent \indent  $X^+(\tau_i+\Delta_i) := X^+(\tau_i) \bigcup T_i'$\\
      \indent \indent {\bf end if} \\
\\

       \indent \indent  $F_{i+1} := \{\phi_q(\xx,\Delta_i)\,:~ \xx \in F_i'  \} \backslash X^+(\tau_i)$~~\\

      \indent{\bf end if} \\
\\

     \indent $i := i+1$~~\\
\\ 

    ~~{\bf end while}~~\\
 
\\
\hline
  \end{tabular}
\\
\\

  As with the previous algorithm, the initialization step ``$F_0 := S_1^+$''
  could be replaced with ``$F_0 := S_0^+$'', depending on convenience.
  Termination of this procedure may be guaranteed, if certain assumptions
  are satisfied. The required assumptions are as follows:
  \begin{enumerate}
   
     \item the set $Y_q = \overline{X_q}$ is compact;
     \item $X_0$ is closed in $\reals^n$; and
     \item for every $\xx_0 \in X_0$, $Y_q$ does not
           contain any $\omega$-limit points of the
           trajectory of $\phi_q(\xx_0,t)$, $t \geq 0$,
           more precisely,
           \[
             \phi_q(\xx_0, \infty) \bigcap Y_q = \emptyset, ~~ \xx_0 \in X_0.
           \]
  \end{enumerate}
  Under these assumptions, it can be shown that (see Appendix C) 
  there is a $\tau_{\max} > 0$, such that for every $\xx \in X_0$,
  there is a $\tau(\xx) > 0$ with $0 < \tau(\xx) \leq \tau_{max}$ and
  $\phi_q(\xx,\tau(\xx)) \not \in X_q$. Obviously, with such a
  $\tau_{\max}$, we have $\tau_{\xx} \leq \tau_{\max}$, for each
  $\xx \in X_0$, where $\tau_{\xx}$ is as in (\ref{taux_inf}).
  Termination of the algorithm may be deduced from this as follows:
  Consider a sequence of sets $F_0^{(i)}$, $i \geq 0$, defined as
  $F_0^{(i)} = \{ \xx \in S_0 : \phi_q(\xx,\tau_i) \in F_i\}$.
  It is easy to check that $F_0^{(i)}$ is a nonincreasing sequence
  of sets, i.e., $F_0^{(i+1)} \subseteq F_0^{(i)}$.  Let $\xx \in X_0$,
  and let $k$ be a nonnegative integer such that
  $\tau_k \leq \tau_{\xx} < \tau_{k+1}$. Now, during the
  $k$'th iteration of the while-loop, the global invariance
  condition (the condition that $T_k \subseteq X_q$) is
  violated, since $\phi_q(\xx, t) \not \in X_q$, for some $t$
  such that $\tau_{\xx} \leq t < \tau_{k+1}$. Let
  $\tau_{\xx}' \in [\tau_{\xx}, \tau_{k+1})$ be a time
  instant such that $\phi_k(\xx,\tau_{\xx}') \not \in X_0$, and let
  $t_{\xx} = \tau_{\xx}'-\tau_k$. With this choice of $t_{\xx}$,
  we have both $0 \leq t_{\xx} < \Delta_k$ and
  $\phi_q(\xx, \tau_k+t_{\xx}) \not \in X_q$. Thus,
  $\phi_q(\xx, \tau_k+t_{\xx}) \in U_k = T_k \backslash X_q$
  (refer to the schematic above). Let $\yy = \phi_q(\xx, \tau_k+t_{\xx})$,
  so $\phi_q(\xx,\tau_k) = \psi_q(\yy,t_{\xx})$. But, 
  since $\yy \in U_k$, $\psi_q(\yy, t_{\xx}) \in V_k$
  (refer to the schematic). Hence $\phi_q(\xx,\tau_k) \in V_k$ and
  $\phi_q(\xx, \tau_k) \not \in F_k' = F_k \backslash V_k$. Therefore
  $\phi_q(\xx, \tau_{k+1}) \not \in F_{k+1}$, implying that
  $\xx \not \in F_0^{(k+1)}$. By the monotonicity of the sequence
  of sets $F_0^{(i)}$, we have $\xx \not \in F_0^{(i)}$, for
  any $i \geq k+1$. Finally, let $m \geq 0$ be an integer
  such that $\tau_m \leq \tau_{\max} < \tau_{m+1}$. Now since
  $\tau_{\xx} \leq \tau_{\max}$, $\xx \not \in F_0^{(m+1)}$,
  for any $\xx \in X_0$. Hence $F_0^{(m+1)} = \emptyset$, so
  $F_{m+1} = \emptyset$, satisfying the terminating condition
  of the while-loop of the algorithm, after at most $m$ number
  of iterations.
\\

  In the rest of the paper, we discuss approximation of the set
  $T_i$ appearing in the second line of the while-loop of the
  above schematic (see also the schematic for $Reach(X_0,[0,\tau])$),
  when $f(q,\xx)$ is of the form $f(q,\xx) = A_q\xx$, where $A_q$ is a
  constant $n\times n$ matrix with real entries and the initial set
  $X_0$ is a polyhedron. For convenience, we drop the subscript $q$
  in $A_q$, and deal with the linear system
  \[
     \frac{d\xx(t)}{dt} = A\xx,
  \]
  where $A$ is a constant $n\times n$ matrix with real valued entries.
\\

 \section {Representation of the Reach Sets}

  \subsection{Preliminary Discussion}

  In this section, we discuss representation of the reach
 set $Reach_q(X_0, [0,\tau])$, such that boolean set operations
 such as union and complementation can be performed efficiently.
 The most convenient representation schemes appear to be representation
 in terms of the following classes of subsets of $\reals^n$: 
  \begin{enumerate}
   \item {\em Polyhedral Sets} :  These are the sets which can be
         written as union of finitely many polyhedra.  A polyhedron in
         $\reals^n$ is a set which may be expressed as the intersection
         of finitely many closed half spaces \cite{Rock70}, i.e.,
         a finite intersection of sets of the form
         $\{\xx \in \reals^n \s: \aa^T\xx \leq b\}$, where
          $\aa \in \reals^n$ and $b \in \reals$. It may be observed
          this class of sets corresponds to (and includes) the class
          of definable sets in  the theory of linear inequalities
          of $\reals$, i.e., the theory obtained when $0$ and $1$ are
          the constant elements, $+$ and $-$ are the (binary) function
          symbols, and $<$ is the (binary) relation symbol (see Appendix A).

 \item {\em Subalgebraic Sets} : These are the sets which can be
         written as union of finite number of sets defined by
         polynomial inequalities, i.e., sets that can be expressed
         as union of finitely many sets, each of which is the
         intersection of finitely many sets the form
         $\{\xx \in \reals^n : p(\xx) \leq c\}$, where
         $p : \reals^n \rightarrow \reals$ is a polynomial function
         with real-valued coefficients and $c \in \reals$. This class
         of sets corresponds to (and includes) the class of sets that
         are definable in the theory of $\reals$ when viewed as an
         ordered field. The theory of ordered field of $\reals$ is
         the theory obtained by extending the theory of linear
         inequalities of $\reals$ by including a binary function
         symbol for representing the product of two real numbers,
         denoted by $\cdot$ (see Appendix A).
         
  \end{enumerate}

  It may be observed that, for a general flow function $\phi_q$,
  the reach set $Reach_q(X_0,[0,\tau])$ may not be exactly
  representable in any one of these classes, even if $X_0$ is. Hence,
  as an alternative, we may have to settle for an approximation of the
  reach set by sets that belong to the respective classes. In this context,
  we shall restrict our attention to over-approximation of the reach set
  with polyhedral sets.
\\

  Specifically, we describe an algorithm for  reach set over-approximation 
 with polyhedral sets, when the system dynamics (within a discrete
 state $q$) are specified as  
\begin{equation}
 \frac{d\xx}{dt} = A\xx, ~~ \xx(0) \in X_0,
\end{equation}
 where the initial set $X_0$ is a polyhedron and $A$ is a constant
 real valued $n \times n$ matrix.  Based on the discussion of the
 last section, we consider the problem of approximating the reach
 set of the solutions of the equation
 \begin{equation}
 \frac{d\xx}{dt} = A\xx, ~~ \xx(0) \in F_0,  \label{LinSys}
\end{equation}   
 where $F_0$ is a face of $X_0$.
%
% We assume that the unit outward
% normal $\lambda$ of $F_0$ exists and is such that, for some $\delta > 0$, 
% $\lambda^T A \xx_0 \geq \delta$, for every $\xx_0 \in F_0$.
%
% We first describe an abstract algorithm, which will be modified in the
% next section in order to present this algorithm in a practicable
% form. This modification requires a further assumption that
% $\lambda^T A \xx_0 \geq \delta > 0$, for $\xx_0 \in F_0$.
\\

 The main references on approximate reach set computation using
 polyhedral sets appear to be \cite{V98} and \cite{ABDM00}. In
 \cite{V98}, the initial set is assumed to be convex (not
 necessarily a polyhedron), and for each time instance, each slice
 of the ``reach tube'' (i.e., for each $t \in [\tau_i, \tau_{i+1}]$,
 the set $\phi(F_i,t)$) is approximated by polyhedra. The method
 described in \cite{ABDM00} assumes $F_i$ to be a polyhedron, and
 approximates the tube for the time interval $[\tau_i, \tau_{i+1}]$,
 i.e., the set $T_i = \bigcup_{0\leq t \leq\tau_{i+1}-\tau_i} \phi(F_i, t)$,
 by first constructing a polyhedral over-approximation of the reach
 tube for the time interval and further over-approximating the
 resulting polyhedron by a ``griddy'' polyhedron, i.e., a set
 that may be expressed as a union of unit hypercubes with integer
 left-most vertices \cite{ABDM00}.
\\

  Being a convex set, any polyhedron that over-approximates the
  reach set contains the convex hull of $T_i$. An algorithm
  based on this observation (as given in \cite{ABDM00}) is to find
  a ``bloated'' convex hull of $F_i$ and $F_{i+1}$. This method
  seems to be simple and straightforward, and gives reasonable
  approximations in a rather short time (refer to \cite{ABDM00}).
  But we may note that those faces of the convex hull of $F_i$
  and $F_{i+1}$, for which the solution set $T_i$ lies on one
  side, need not be bloated. By shrinking the convex hull,
  an under-approximation is obtained by the same method.
\\

  If the time steps are constant, i.e., $\tau_k = k\tau$, for
  some $\tau > 0$, as in \cite{ABDM00}, then we have to find
  approximation $P_0$ only for $T_0$, since symbolically,
  $T_k = e^{(A\tau)}T_{k-1}$, for $k \geq 1$, and therefore,
  $T_k = e^{(kA\tau)}T_0 = e^{(A\tau_k)}T_0$. Thus,
  $P_k =  e^{(A\tau_k)}P_0$ is the required approximation, and
  if $\lambda_j$, $j = 1,2,\ldots, N$, are the normals of the
  faces $P_0$, then the normals of the faces of $P_k$ are given
  by $e^{(-A^T\tau_k)} \lambda_j$.  From this, we may infer that
  in the subsequent iterations of the algorithm of \cite{ABDM00},
  the convex hull of $F_i$ and $F_{i+1}$ need not be computed.

  \subsection{Over-approximating the Reach Set with Polyhedra}

  In this section, we consider the problem of over-approximating the
  reach sets, when the dynamics are specified as
  \[
    \frac{d\xx(t)}{dt} = A \xx(t),   ~ t \geq 0,
  \]
  with the initial conditions $\xx(0) \in X_0$. The solution
  is given explicitly by $\xx(t) = e^{(At)}\xx(0)$. We assume
  that the initial set $X_0$ is a polyhedron, defined by
  \begin{equation}
        \lambda_i^T \xx - h_i \leq 0, ~~ i = 1, 2, \ldots, \kappa,
  \end{equation}
  where $\kappa$ is a fixed positive integer. We assume that
  $\lambda_i \in \reals^n$, $|\lambda_i | = 1$ and $h_i \in \reals$, for
  $1 \leq i \leq \kappa$. Then the boundary set of $X_0$ consists of
  sets $C_j$, $1 \leq j \leq \kappa$, of the form 
  \begin{eqnarray*}
        \lambda_j^T \xx - h_j & = &  0, ~~ \textrm {and} \\
        \lambda_i^T \xx - h_i & \leq &  0, ~~ 
                 \textrm{ for  $1 \leq i \leq \kappa $  and  $ i \neq j$ }.
  \end{eqnarray*}
 Further, we consider only those polyhedral initial sets $X_0$ and
 matrices $A$ that satisfy the following assumptions:
   \begin{enumerate}
     \item[($\acal_1$)]  $X_0$ is compact.
     \item[($\acal_2$)] The set $S_0^+$, as defined in (\ref{S_0_plus})
               corresponding to the flow function
               $\phi_q(\xx,t) = e^{(At)}\xx$, can be expressed as
               $S_0^+ = C_{j_1} \bigcup C_{j_2} \bigcup \ldots \bigcup C_{j_m}$,
               where $1 \leq j_1 < j_2 < \ldots < j_m \leq \kappa$, such that
               for each $i$ with $1 \leq i \leq m$, 
               $\lambda_{j_i}^T A \xx - h_{j_i} \geq \delta_i > 0$, for
               every $\xx \in C_{j_i}$.
   \end{enumerate}
   Referring to these assumptions, it may be mentioned that the
   main limitation of this method appears to be the restriction
   imposed by assumption $\acal_2$ on the initial set $X_0$ and the
   matrix $A$. In contrast, the method described in \cite{ABDM00}
   does not require such an assumption. But for our purposes, we
   need this assumption.
\\

  Now, let $j$ be an index such that $C_j \subseteq S_0^+$,
  as in assumption ($\acal_2$). For simplicity and definiteness,
  let us assume that $j = \kappa$ and put $F_0 = C_{\kappa}$.
  Hence, $F_0$
  is described by
  \begin{eqnarray*}    
        \lambda_i^T \xx - h_i & \leq &  0, ~~ 
                 \textrm{ $1 \leq i \leq \kappa-1$, ~ and }\\
        \lambda_{\kappa}^T \xx - h_{\kappa} & = & 0.
  \end{eqnarray*}
  After some manipulations and rearrangements, if necessary,
  the above system of linear constraints can be transformed into an
  equivalent system of linear constraints, which also describes $F_0$
  but is of the following form:
  \begin{equation}    
   \left.
      \begin{array}{lll}
        \aa_i^T \xx - b_i & \leq &  0, ~~ 
                 \textrm{ $1 \leq i \leq k-1$, ~ and }\\
        \aa_k^T \xx - b_k & = & 0,
      \end{array}
       \right \}  \label{ConstraintsForF0}
   \end{equation}
  where $k \leq \kappa$, $\aa_k = \lambda_{\kappa}$, and for
  $1 \leq i < k$, $|\aa_i | = 1$ and $\aa_i^T \aa_k = 0$. It may
  be observed that, since $|\lambda_k| = 1$ initially, we have
  $|\aa_k| = 1$ as well. Also, by assumption ($\acal_2$), there
  is a $\delta > 0$, such that for all the vectors $\xx \in F_0$,
  $\aa_k^T A\xx \geq \delta$. Further, we assume that the system
  of linear constraints in (\ref{ConstraintsForF0}) is consistent
  and that each of the constraints is linearly independent from
  the remaining constraints.
 \\

  With this construction, our objective is to describe an algorithm
  to find a polyhedron $P_0$ as an over-approximation to the set
  \begin{equation}
     T_0 = \{e^{(At)}\xx_0\s : ~ \xx_0 \in F_0, ~ t \in [0,\Delta]\},
        \label{T0_set}
  \end{equation}
  where $\Delta$ is a small positive number for which the 
  the following condition holds:
  \begin{enumerate}
     \item[($C_1$)] for some $\delta_0 > 0$,
            $\aa_k^T A e^{(At)} \xx_0 \geq \delta_0$,
           for all $\xx_0 \in F_0$ and for all $t \in [-\Delta,\Delta]$.
  \end{enumerate}
  The existence of such a $\Delta$ can be assured by our
  assumptions ($\acal_1$) and ($\acal_2$). Later, we shall
  derive an estimate for such a $\Delta$, for any fixed
  $\delta_0 > 0$ such that $\delta_0 < \delta$. In comparison,
  the method of \cite{ABDM00} does not impose any such
  restrictions on $\Delta$. But, it may be mentioned that,
  in both cases, the accuracy of approximation of either
  method depends on how small a value is chosen for the
  parameter $\Delta$. The two conditions ($C_2$) and ($C_3$)
  stated below follow from condition ($C_1$):
  \begin{enumerate}
     \item[($C_2$)] $\aa_k^T e^{(At)} \xx_0 - b_k > 0$,
            for all $\xx_0 \in F_0$ and for all $t$ such that $0 < t \leq \Delta$; and 
     \item[($C_3$)] $\aa_k^T e^{(At)} \xx_0 - b_k < 0$,
            for all $\xx_0 \in F_0$ and for all $t$ such that $-\Delta \leq t < 0$.
  \end{enumerate}

   Before proceeding to describe our algorithm, we note that the
   set $F_{\Delta} = \{ e^{(A\Delta)}\xx_0\s:~ \xx_0 \in F_0 \}$ may be
   described by the system of linear constraints
   \begin{eqnarray*}
        \aa_i^T(\Delta) \xx - b_i & \leq & 0,
                                         ~~ i = 1, 2, \ldots, k-1, \\
        \aa_k^T(\Delta) \xx - b_k & = & 0.
   \end{eqnarray*}
   where $\aa_i(\Delta) = e^{(-A^T\Delta)}\aa_i$, $1 \leq i \leq k$.
   Dividing throughout the last equation in the above system
   by $|\aa_k(\Delta)|$, we get
   \begin{eqnarray*}
        \aa_i^T(\Delta) \xx - b_i & \leq & 0,
                                         ~~ i = 1, 2, \ldots, k-1, \\
        \bb_k^T \xx - b_k' & = & 0,
   \end{eqnarray*}
   where $\bb_k = \frac{\aa_k(\Delta)}{|\aa_k(\Delta)|}$ and 
   $b_k' = \frac{b_k}{|\aa_k(\Delta)|}$.  Now, we observe that,
   since $e^{(-A^T\Delta)}$ is invertible and $|\aa_i| = 1$, each
   $\aa_i(\Delta)$ is nonzero, and since $\aa_k$ is orthogonal to
   $\aa_i$, for $1 \leq i \leq k-1$, $\aa_k(\Delta)$ and $\aa_i(\Delta)$
   are pair-wise independent.  So, after subtracting from each of the
   inequalities of the above system an appropriate constant times the
   last equation, and after normalization, the above system of linear
   constraints can be transformed into the following system of linear
   constraints:
   \begin{equation}
    \left.
     \begin{array}{lll}
        \bb_i^T \xx - b_i' & \leq & 0,
                                         ~~ i = 1, 2, \ldots, k-1, \\
        \bb_k^T \xx - b_k' & = & 0,
     \end{array}
      \right \}  \label{ConstraintsForF-Delta}
   \end{equation}
   where, for $1 \leq i \leq k$, $|\bb_i| = 1$, and for
   $1 \leq i \leq k-1$, $\bb_i^T\bb_k = 0$.
   Now, by the condition ($C_3$), for $0\leq t < \Delta$ and $\xx_0 \in F_0$,
   $\aa_k^T(\Delta) e^{(At)}- b_k = \aa_k^T e^{\left(A(t-\Delta)\right)}-b_k < 0$.
   Hence, we also have, for $0\leq t < \Delta$ and $\xx_0 \in F_0$,
   $\bb_k^T e^{(At)}\xx_0 - b_k' <  0$.
\\

%   Now, we observe that, since $e^{(-A^T\Delta)}$ is invertible and
%   $|\aa_i| = 1$, each $\aa_i(\Delta)$ is nonzero, and since $\aa_k$
%   is orthogonal to $\aa_i$, for $1 \leq i \leq k-1$, $\aa_k(\Delta)$
%   is independent from each of the vectors $\aa_i(\Delta)$,
%   $1 \leq i \leq k-1$. So the vector
%   $\aa_i'(\Delta) = \aa_i(\Delta)-\alpha_i\aa_k(\Delta)$, where
%   $\alpha_i = \frac{\aa_k(\Delta)^T\aa_i(\Delta)}{|\aa_k(\Delta)|^2}$,
%   is nonzero, for $1 \leq i \leq k-1$, and orthogonal to $\aa_k(\Delta)$.  
%   Let $\bb_k = \frac{\aa_k(\Delta)}{|\aa_k(\Delta)|}$; and for
%   $1 \leq i \leq k-1$, let $\bb_i = \frac{\aa_i'(\Delta)}{|\aa_i'(\Delta)|}$;
%   also, let $b'_k = \frac{b_k}{|\aa_k(\Delta)|}$ and for $1 \leq i \leq k-1$,
%   let $b_i' = \frac{b_i-\alpha_i b_k}{|\aa_i'(\Delta)|}$. With these
%   transformations, we now can express $F_{\Delta}$ by the system of
%   linear constraints
%   \begin{eqnarray*}
%        \bb_i^T \xx - b_i' & \leq & 0,
%                                         ~~ i = 1, 2, \ldots, k-1, \\
%        \bb_k^T \xx - b_k' & = & 0,
%   \end{eqnarray*}
%   where, and for $1 \leq i \leq k$, $|\bb_i| = 1$,
%   and for $1 \leq i \leq k-1$, $\bb_i^T\bb_k = 0$.
%   Now, by the condition ($C_3$), for $0\leq t < \Delta$ and $\xx_0 \in F_0$,
%   $\aa_k^T(\Delta) e^{(At)}- b_k = \aa_k^T e^{\left(A(t-\Delta)\right)}-b_k < 0$.
%   Hence, we also have, for $0\leq t < \Delta$ and $\xx_0 \in F_0$,
%   $\bb_k^T e^{(At)}\xx_0 - b_k' <  0$.
%\\
%
       
 In this notation, we now describe a schematic algorithm, the output
 of which is a set of $4k$ many parameters,  consisting of pairs of
 vectors and constants, $(\mu_i,c_i)$ and $(\nu_i, d_i)$,
 $i = 1, 2, \ldots, 2k$, where for each $i$, $\mu_i, \nu_i \in \reals^n$
 and $c_i, d_i \in \reals$, such that the polyhedron $P_0$ of
 intersection of the $4k$ half-spaces defined by
 $L_i = \{\xx \in \reals^n \s: ~ \mu_i^T\xx - c_i \leq 0\}$ and 
 $L_i' = \{\xx \in \reals^n \s: ~ \nu_i^T\xx - d_i \leq 0\}$ is
 an over-approximating polyhedron for the set $T_0$ as defined in
 (\ref{T0_set}).
\\

%\newpage
  \begin{tabular}{|l|}
  \hline
\\
   ~~~~~ {\sf   Schematic Algorithm for Finding Over-approximating Polyhedron} \\
\\
  {\bf Reach Set :} $T_0 = \{e^{(At)}\xx_0 : ~ t \in [0,\Delta], ~\xx_0 \in F_0 \}$
                  (refer to (\ref{T0_set}))\\

  {\bf Output :} ~~{\sf $4k$ vectors $\mu_i$, $\nu_i$ and $4k$ real numbers $c_i$, $d_i$, $1 \leq i \leq 2k$  }\\
\\

\indent  /* ~  {\sf the $4k$ half-spaces are $L_i$ and $L_i'$, $1\leq i \leq 2k$, \s where} ~ */\\
\indent  /* ~ {\sf $L_i = \{\xx \in \reals^n \s: ~ \eta_i (\xx) = \mu_i^T\xx - c_i \leq 0\}$, ~ and } ~~~~~~~ */\\
\indent  /* ~ {\sf $L_i' = \{\xx \in \reals^n \s: ~ \eta_i' (\xx) = \nu_i^T\xx - d_i \leq 0\}$ } ~~~~~~~~~~~~~~~ */ \\
\\

  {\bf  for}  $i = 1, 2, \ldots, k-1$,  {\bf do} ~~ /* to find ($\mu_i$, $c_i$) and ($\nu_i$, $d_i$) */\\

       \indent   $l_i := \inf \{ l\s: (\aa_i^T\xx-b_i)-l(\aa_k^T\xx-b_k) \leq 0,$
             $\forall \xx \in T_0 \}$; \\

\indent          $\mu_i := \aa_i-l_i\aa_k$; ~~ $c_i := b_i-l_i b_k$;\\
\\
 
\indent          $l_i' := \inf \{ l'\s: (\bb_i^T\xx-b_i')+l'(\bb_k^T\xx-b_k')
\leq 0,$
               $\forall \xx \in T_0 \}$;\\
 
       \indent    $\nu_i := \bb_i+l_i'\bb_k$; ~~ $d_i = b_i'+l_i'b_k'$;\\

  {\bf end for} \\

          $\mu_k := -\aa_k^T$;~~ $c_k = - b_k$;\\
          $\nu_k := \bb_k^T$; ~~ $d_k = b_k'$;\\
\\

              $l_{k} := \inf \{ l\s: (\aa_k^T\xx-b_k)-l \leq 0,$ 
               $\forall \xx \in T_0 \}$; \\

              $\mu_{k} := \aa_k$; ~~  $c_{k+1} := b_k+l_{k}$;\\
\\

              $l_{k}' := \inf \{ l\s: -(\bb_k^T\xx-b_k')-l \leq 0$, 
              $\forall \xx \in T_0 \}$; \\

              $\nu_{k} := -\bb_k$; ~~  $d_{k+1} := -b_k'+l_{k}'$;\\

\\
%\\
%  ~~ /* the following $2k-2$ parameters are for additional half-spaces */\\
%\\

  {\bf for}  $i = 1, 2, \ldots, k-1$, {\bf do}   \\
 \indent \indent /* to find ($\mu_{i+k+1}, c_{i+k+1}$) and ($\nu_{i+k+1}, d_{i+k+1}$) */\\

       \indent  $l_{i+k} = \inf \{ l\s: (\aa_i^T\xx-b_i)-l \leq 0,$
             $\forall \xx \in T_0 \}$;  \\

       \indent $\mu_{i+k} = \aa_i$; ~~ $c_{i+k+1} = b_i+l_{i+k}$;\\
 
 \\
 
         \indent $l_{i+k}' = \inf \{ l'\s: (\bb_i^T\xx-b_i')-l' \leq 0,$
               $\forall \xx \in F_0 \}$;\\

         \indent $\nu_{i+k} = \bb_i$; ~~ $d_{i+k+1} = b_i'+l_{i+k}'$;\\

  {\bf end for} \\
 \\

  \hline

 \end{tabular}
\\

 We first observe that both $l_i > -\infty$ and $l_i' > -\infty$.
% since $T_0$ is a compact set.
 If $\aa_k^TA\xx \geq \delta > 0$,
 as in assumption $\acal_2$, then both $l_i < \infty$ and $l_i' < \infty$,
 as we shall see in Sec. \ref{UpperBoundsForL_i} below. If $\Delta$
 is small, such that the set $T_0$ is nearly a polyhedron, then
 the above method gives reasonable results. Referring to Step 1
 of the first for-loop of the above algorithm, the reason for
 choosing $l_i$ to be the infimum over all $l$ for which
 $(\aa_i^T\xx-b_i)-l(\aa_k^T\xx-b_k) \leq 0$,
 $\forall \xx \in T_0$, is obvious: if
 $P_1$ and $P_2$ are two polyhedra such that $P_1$ is obtained by
 the above algorithm and $P_2$ is obtained by replacing a constraint
 $\mu_i^T\xx-c_i \leq 0$, where $\mu_i$ and $c_i$ are as in Step 2,
 with another constraint $\lambda_i^T\xx-h_i \leq 0$, where
 $\lambda_i = \aa_i-l\s \aa_k$ and $h_i = b_i-l\s b_k$, for some $l > l_i$,
 then $P_1 \subset P_2$.\footnote{This follows from the observation
 that if both $(\aa_i^T\xx-b_i)-l(\aa_k^T\xx-b_k) \leq 0$
 and $-(\aa_k^T\xx-b_k) = \mu_k^T \xx - c_k \leq 0$, where $l \in \reals$ and
 $\xx \in \reals^n$, $\xx$ not necessarily in $T_0$, then, for any
 $h > l$, $(\aa_i^T\xx-b_i)-h(\aa_k^T\xx-b_k) \leq 0$.}
 (An analogous statement holds for each of the remaining parameters, $l_i'$,
 $l_k$, $l_k'$, $l_{k+i}$ and $l_{k+i}'$, as in the algorithm.)
\\

 It can be shown that the polyhedron included in the $k+1$ half-spaces,
 specified by $L_1, L_2, \ldots, L_k, L_{k+1}$, is a bounded polyhedron
 (see Appendix D). We may also note that the hyperplanes
 $\mu_i^T \xx -c_i = 0$ and $\nu_i^T \xx -d_i = 0$, $1 \leq i \leq k-1$,
 are obtained by rotating the hyperplanes $\aa_i^T \xx -b_i = 0$ and
 $\bb_i^T\xx - b_i' = 0$ about their corresponding intersection with
 the hyperplanes $\aa_k^T \xx - b_k= 0$ and $\bb_k^T \xx - b_k' = 0$,
 respectively; whereas the hyperplanes $\mu_{k+1}^T\xx -c_{k+1} = 0$
 and $\nu_{k+1}^T\xx-d_{k+1} = 0$ are obtained by translating the
 hyperplanes $\aa_k^T\xx-b_k = 0$ and $\bb_k^T\xx-b_k = 0$, respectively.
 Similarly, for $i = 1, 2, \ldots, k-1$, the hyperplanes
 $\mu_{k+i}^T \xx -c_{k+i} = 0$ and $\nu_{k+i}^T \xx -d_{k+i} = 0$
 are translations of the hyperplanes $\aa_i^T \xx - b_i = 0$ and
 $\bb_i^T \xx - b_i' = 0$, respectively.
\\

 In remaining part of the section (in Sec. \ref{UpperBoundsForL_i} below),
 we shall derive upper bounds for the numbers $l_i$ and $l_i'$,
 $1 \leq i \leq 2k-1$, that are defined in the schematic. If all
 the $l_i$ and $l_i'$ are replaced with their corresponding upper bounds,
 $\hat{l}_i$ and $\hat{l}_i'$ respectively, then we obtain conservative
 estimates -- $\hat{\mu}_i$, $\hat{c}_i$, $\hat{\nu}_i$ and $\hat{d}_i$
 respectively -- for $\mu_i$, $c_i$, $\nu_i$ and $d_i$.  After this
 replacement, we obtain a modified algorithm for over-approximation
 with polyhedra. It may be observed that these estimates overcome the
 difficulties in finding the infima required for obtaining $l_i$'s and
 $l_i'$'s. But unfortunately, the conservative upper bounds that we
 derive for $l_i$'s and $l_i'$'s (as in Sec. \ref{UpperBoundsForL_i})
 may turn out to be very large. However, better accuracy may be
 obtained if the modified alogrithm is used in conjunction with that
 of \cite{ABDM00} (see Fig. \ref{FigPoly3} in Sec. 5). More precisely,
 the intersection of the polyhedron obtained by the method described
 here with that obtained by the method of \cite{ABDM00} gives a smaller
 over-approximating polyhedron, as will be discussed in Sec. 5, while
 illustrating these algorithms with simple examples.

\subsection{\label{UpperBoundsForL_i} Upper Bounds for $l_i$ and $l_i'$} 

  In this section, we derive upper bounds for the numbers $l_i$ and $l_i'$
  appearing in the schematic algorithm. We begin with the following result:

  \begin{enumerate}

    \item[] {\bf Claim 1.} For $i = 1, 2, \ldots, k$, with $\aa_i$ and $b_i$
                       as in (\ref{ConstraintsForF0}) and $\xx_0 \in F_0$ and
                         $t \in (0, \Delta]$, we have
                     \[
                     \left |\frac{ (\aa_i^T e^{(At)} \xx_0- \aa_i^T \xx_0) } { t } \right | \leq
                                M_0 \|\aa_i^TA\| e^{(\|A\|\Delta)},
                     \]
                     where $M_0 = \max_{\xx_0 \in F_0} \{ \|\xx_0\| \}$.
  \end{enumerate}

 To prove the claim, we first note that $(\aa_i^Te^{(At)}\xx_0-\aa_i^T\xx_0)$
  $ = t\times \aa_i^T A e^{(A\theta)}\xx_0$,  for some $\theta$ with
  $0 < \theta < t$, where $\theta$ may depend on $t$. Hence, we have
 \begin{eqnarray*}
\left |  \frac{ (\aa_i^T ee^{(At)} \xx_0-\aa_i^T \xx_0) } { t }  \right |
          & =  &  | \aa_i^T A e^{(A\theta)} \xx_0|, ~~
                        \textrm{where $\theta$ is such that $0 < \theta < t$} \\
          & \leq & |\aa_i^T A e^{(A\theta)}| \times \|\xx_0\|\\
          & \leq & M_0 \|\aa_i^TA\| e^{(\|A\|\Delta)}.
 \end{eqnarray*}
 
\noindent With $\bb_i$ and $b_i'$, we have an analogous result, but with 
 a slight modification, as in the following:
  \begin{enumerate}
 
    \item[] {\bf Claim 2.} For $i = 1, 2, \ldots, k$, with $\bb_i$ and $b_i'$
                      as in (\ref{ConstraintsForF0}) and $\xx_0 \in F_0$ and
                      $t \in [0, \Delta)$, we have
                     \[
    \left | \frac{ (\bb_i^T e^{(At)} \xx_0-\bb_i^T e^{(A\Delta)} \xx_0) }{(\Delta - t)} \right | \leq
                                M_0  \|\bb_i^T A\| e^{(\|A\|\Delta)},
                     \]
         where $M_0 = \max_{\xx_0 \in F_0} \{ \|\xx_0\| \}$. 
  \end{enumerate}
  We have $(\bb_i^T e^{(At)} \xx_0-\bb_i^T e^{(A\Delta)} \xx_0)$
  $= (t-\Delta) \times \bb_i^T A e^{(A\theta)} \xx_0$, for some $\theta$ with
  $t < \theta < \Delta$. Hence
 \begin{eqnarray*}
 \left | \frac{ (\bb_i^T e^{(At)} \xx_0-\bb_i^T e^{(A\Delta)} \xx_0) }{(\Delta - t)} \right |
    & = & | \bb_i A e^{(A\theta)} \xx_0|,
          ~~ \textrm{ for some $\theta$ with $t < \theta < \Delta$ } \\
    & \leq &  \| \bb_i^T A e^{(A\theta)}\| \times \| \xx_0 \|\\
         & \leq & M_0 \|\bb_i^TA\| e^{(\|A\|\Delta)}.
  \end{eqnarray*} 

  Finally, we shall make the following claim, before deriving upper bounds for
  $l_i$'s and $l_i'$'s:
  \begin{enumerate}
     \item[] {\bf Claim 3.} Assume, for some $\delta > 0$,
     $\aa_k^T A \xx_0 \geq \delta$, for every $\xx_0 \in F_0$, and
     let $\delta_0$ be such that $0 < \delta_0 < \delta$. Then there
     is a $\Delta > 0$ such that $\aa_k^T A e^{(At)} \xx_0 \geq \delta_0$,
     for all $\xx_0 \in F_0$ and for all $t \in [-\Delta,\Delta]$.
  \end{enumerate}                                                                  
  In order to prove the claim, we shall derive a conservative estimate
  for a $\Delta > 0$ for which the claim holds. Fix a $\delta_0$ such that
  $0 < \delta_0 < \delta$. We have
  \vspace*{-0.1in}
 \begin{eqnarray*}
   \aa_k^T A e^{(At)} \xx_0
      & = & \aa_k^T A \xx_0 + \int_0^t \aa_k^T A^2 e^{(A s)} \xx_0\,{\rm d}s,
                ~~~ \textrm{ where $\xx_0 \in F_0$,} \\
      & \geq & \aa_k^T A\xx_0 -
           \int_0^{|t|} \|\aa_k^T\| \|A\|^2 e^{(\|A\|s)}\|\xx_0\| \,{\rm d}s\\
      & \geq & \aa_k^T A\xx_0 - M_0 \|A\| (e^{\|A\| |t|} - 1),
  \end{eqnarray*}
  where $M_0 = \max_{\xx_0 \in F_0} \{ \|\xx_0\| \}$. Therefore, if
  $\Delta$ is such that
  \[
     M_0 \|A\| (e^{\|A\| \Delta} - 1) \leq \delta - \delta_0,
  \]
  then the claim holds. 

\paragraph{Upper Bounds for $l_i$ and $l_i'$, $1 \leq i \leq k-1$.}
 Recall that, for $1 \leq i \leq k-1$,
  \begin{equation}
   l_i = \inf
 \{ l\s: (\aa_i^T\xx-b_i)-l(\aa_k^T\xx-b_k) \leq 0,
               ~\forall  \xx \in T_0 \}.  \label{L_i}
  \end{equation}
  Referring to the right hand side of (\ref{L_i}), for every
  $\xx_0 \in F_0$, since $\aa_k^T\xx_0-b_k = 0$, for any real
  number $l$, we have
  $(\aa_i^T\xx_0-b_i)-l(\aa_k^T\xx_0-b_k) = (\aa_i^T\xx_0-b_i) \leq 0$.
  Therefore, we may assume $\xx \in T_0 \backslash F_0$, and
  $(\aa_k^T\xx-b_k) > 0$. Thus $l_i$ has to be chosen such that
  \[
   l_i = \sup_{\xx \in T_0 \backslash F_0 }
              \frac{(\aa_i^T\xx-b_i)}{(\aa_k^T\xx-b_k)}.
  \]
  Rewriting the above, we have
  \[
    l_i = \mathop {\sup_{0 < t \leq \Delta}}_{\xx_0 \in F_0}
              \frac{(\aa_i^Te^{(At)}\xx_0-b_i)}{(\aa_k^Te^{(At)}\xx_0-b_k)}.
  \]                                                                               
  Dividing the numerator and denominator by $t$, this may be
  written as
   \[
    l_i = \mathop {\sup_{0 < t \leq \Delta}}_{\xx_0 \in F_0} \frac{\alpha_i(\xx_0,t)}{\beta_i(\xx_0,t)},
  \]
  where $\alpha_i(\xx_0,t)$ and $\beta_i(\xx_0,t)$ are given by
  \[
     \alpha_i(\xx_0, t)  =   \frac{(\aa_i^Te^{(At)}\xx_0-b_i)}{t} \textrm{~ and ~}
     \beta_i(\xx_0, t)  = \beta (\xx_0, t) =  \frac{(\aa_k^Te^{(At)}\xx_0-b_k)}{t}.
  \]                                                                               
  But, since $\aa_i^T\xx_0-b_i \leq 0$, for every $\xx_0 \in F_0$, we have
  \[
     \frac{(\aa_i^Te^{(At)}\xx_0-b_i)}{t} ~ \leq ~
               \frac{(\aa_i^Te^{(At)}\xx_0-\aa_i^T\xx_0)}{t} ~  \leq ~
               \left | \frac{(\aa_i^Te^{(At)}\xx_0-\aa_i^T\xx_0)}{t} \right |.
  \]
  By {\bf Claim 1}, we have
 \[
    \mathop {\sup_{0 < t \leq \Delta}}_{\xx_0 \in F_0}  \alpha_i(\xx_0, t) \leq
                                M_0 \|\aa_i^TA\| e^{(\|A\|\Delta)}.
 \]
  As to the denominator, since $\aa_k^T\xx_0-b_k = 0$,
  for every $\xx_0 \in F_0$, we have
  \[
     \frac{(\aa_k^Te^{(At)}\xx_0-b_k)}{t} ~ = ~
               \frac{(\aa_k^Te^{(At)}\xx_0-\aa_k^T\xx_0)}{t} ~  = ~
                \aa_k^T A e^{(A\theta)}\xx_0 ~ \geq ~ \delta_0,
  \]
  where, $\theta$ is some number in the interval $(0,t)$, and the last
  inequality is due to {\bf Claim 3}. Hence
  \[
    \mathop {\inf_{0 < t \leq \Delta}}_{\xx_0 \in F_0} \beta (\xx_0, t) 
          \geq  \delta_0 > 0.
  \]
  Combining both inequalities, we have
  \[
    l_i ~  \leq ~  \frac{ \sup_{0 < t \leq \Delta, ~ \xx_0 \in F_0}
  \alpha_i(\xx_0,t)  }  { \inf_{0 < t \leq \Delta, ~ \xx_0 \in F_0}
    \beta_i(\xx_0,t)  } ~  \leq ~  \hat{l}_i ~  = ~  \frac{ M_0 \|\aa_i^TA\| e^{(\|A\|\Delta)}} {\delta_0}.
  \]
  
  Likewise, $l_i'$ must be chosen such that
 \[
    l_i' = \mathop {\sup_{0 \leq t < \Delta}}_{\xx_0 \in F_0} \frac{\alpha_i'(\xx_0,t)}{\beta_i'(\xx_0,t)},
  \]
  where $\alpha_i'(\xx_0,t)$ and $\beta_i'(\xx_0,t)$ are given by
 \[
  \alpha_i'(\xx_0, t)  =   \frac{(\bb_i^Te^{(At)}\xx_0-b_i')}{(\Delta-t)} \textrm{~ and ~}
 \beta_i'(\xx_0, t)  = \beta' (\xx_0, t) =  \frac{(\bb_k^Te^{(At)}\xx_0-b_k')}{(\Delta-t)}. 
 \]
 Now, since $\bb_i^Te^{(A\Delta)}\xx_0-b_i' \leq 0$, for $\xx_0 \in F_0$,
 \begin{eqnarray*}    
\frac{(\bb_i^Te^{(At)}\xx_0-b_i')}{(\Delta-t)} & \leq & 
  \frac{(\bb_i^Te^{(At)}\xx_0-\bb_i^Te^{(A\Delta)}\xx_0)}{(\Delta-t)} \\
  & \leq & \left |
   \frac{(\bb_i^Te^{(At)}\xx_0-\bb_i^Te^{(A\Delta)})}{(\Delta-t)}\right | \\
         & \leq &  M_0 \|\bb_i^TA\| e^{(\|A\|\Delta)}~, ~~~~\textrm {by {\bf Claim 2}}.
 \end{eqnarray*}
 For the denominator function, since $\bb_k^Te^{(A\Delta)}\xx_0-b_k' = 0$,
 for every $\xx_0 \in F_0$, we have
  \[
     \frac{(\bb_k^Te^{(At)}\xx_0-b_k')}{(\Delta-t)} ~  = ~
          \frac{(\bb_k^Te^{(At)}\xx_0-\bb_k^T e^{(A\Delta)}\xx_0)}{(\Delta-t)}
             ~ = ~ - \bb_k^T A e^{(A\theta)}\xx_0,   
  \]
  where $\theta$ is some number in the interval $(t,\Delta)$. Now,
  \[
  - \bb_k^T A e^{(A\theta)}\xx_0 =
  \frac{ \aa_k^T e^{(-A\Delta)} A e^{(A\theta)} \xx_0} {\|\aa_k^T e^{(-A\Delta)}\|} \geq
      \delta_1 = \frac{\delta_0}{\|\aa_k^T e^{(-A\Delta)}\|}.
\]
 Hence, we have the following upper bound for $l_i'$:
\[
    l_i' ~  \leq ~  \frac{ \sup_{0 < t \leq \Delta, ~ \xx_0 \in F_0}
  \alpha_i'(\xx_0,t)  }  { \inf_{0 < t \leq \Delta, ~ \xx_0 \in F_0}
    \beta_i'(\xx_0,t)  } ~  \leq ~  \hat{l}_i' ~  = ~  \frac{ M_0 \|\aa_i^TA\| e^{(\|A\|\Delta)}} {\delta_1}.
  \]   

\noindent Thus, we obtain the following upper bounds:
\begin{eqnarray*}
  l_i & \leq & \hat{l}_i ~ = ~ \frac {M_0 \|\aa_i^T A\| e^{(\|A\|\Delta)}}{\delta_0},
 ~~ 1 \leq i \leq k-1 \\
  l_i' & \leq & \hat{l}_i' ~ = ~ \frac {M_0 \|\aa_i^T A\| e^{(\|A\|\Delta)}}{\delta_1},
 ~~ 1 \leq i \leq k-1, ~~ \textrm { where } \\
& &      \delta_1 ~ = ~  \frac{\delta_0}{\|\aa_k^T e^{(-A\Delta)}\|}.
\end{eqnarray*}

\paragraph{Upper Bounds for $l_k$, $l_k'$, $l_{k+i}$ and $l_{k+i}'$, $1 \leq i \leq k-1$.}
We have
$l_{i+k} = \inf \{ l\s: (\aa_i^T\xx-b_i)-l \leq 0,~~ \forall \xx \in T_0 \}$,
and $l_{k} = \inf \{ l\s: (\aa_k^T\xx-b_k)-l \leq 0,~~\forall \xx \in T_0 \}$.
Let $l_{2k} = l_k$, so we may consider $l_{k+i}$, for $1 \leq i \leq k$.
We have to choose $l_{k+i}$ such that
\[
   l_{k+i} = \sup_{\xx \in T_0} (\aa_i^T\xx-b_i) = 
     \mathop {\sup_{0 < t \leq \Delta}}_{\xx_0 \in F_0}
              (\aa_i^Te^{(At)}\xx_0-b_i).
\]
Now, since $(\aa_i^T\xx_0-b_i) \leq 0$, for every $\xx_0 \in F_0$, we have
\begin{eqnarray*}
  \mathop {\sup_{0 < t \leq \Delta}}_{\xx_0 \in F_0} (\aa_i^Te^{(At)}\xx_0-b_i)
     & \leq & \mathop {\sup_{0 < t \leq \Delta}}_{\xx_0 \in F_0}
                 (\aa_i^Te^{(At)}\xx_0-\aa_i^T\xx_0)\\
     & \leq & \mathop {\sup_{0 < t \leq \Delta}}_{\xx_0 \in F_0}
                 t \times \frac{ (\aa_i^Te^{(At)}\xx_0-\aa_i^T\xx_0) } {t}\\
     & \leq & \Delta \times \mathop {\sup_{0 < t \leq \Delta}}_{\xx_0 \in F_0}
                 | \frac{ (\aa_i^Te^{(At)}\xx_0-\aa_i^T\xx_0) } {t} |\\
     & \leq & \Delta \times M_0 \|\aa_i^TA\| e^{(\|A\|\Delta)}. 
\end{eqnarray*}

Similarly, for $1 \leq i \leq k-1$, we have to choose $l_{k+i}'$ as follows
\[
   l_{k+i}' = \sup_{\xx \in T_0} (\bb_i^T\xx-b_i') =
     \mathop {\sup_{0 \leq t < \Delta}}_{\xx_0 \in F_0}
              (\bb_i^Te^{(At)}\xx_0-b_i').
\]       
A calculation similar to the above shows 
\[
     \mathop {\sup_{0 \leq t < \Delta}}_{\xx_0 \in F_0}
              (\bb_i^Te^{(At)}\xx_0-b_i') ~ \leq  ~
      \Delta \times M_0 \|\bb_i^T A\| e^{(\|A\|\Delta)}. 
\]
Finally, for $l_k'$, we have
\[
   l_k' = \sup_{\xx \in T_0} [-(\bb_k^T\xx-b_k')]  =
     \mathop {\sup_{0 \leq t < \Delta}}_{\xx_0 \in F_0}
              (\bb_k^Te^{(A\Delta)}\xx_0-\bb_k^Te^{(At)}\xx_0) \\
\]
Another sequence of similar calculations shows
\[
     \mathop {\sup_{0 \leq t < \Delta}}_{\xx_0 \in F_0}
              (\bb_k^Te^{(A\Delta)}\xx_0-\bb_k^Te^{(At)}\xx_0) 
              ~ \leq ~
          \Delta \times M_0 \|\bb_k^T A\| e^{(\|A\|\Delta)}. 
\]
So, to collect all the estimates, we have
\begin{eqnarray*}
    l_k & \leq & \hat{l}_k =  M_0 \Delta \|\aa_k^T A\| e^{(\|A\|\Delta)}\\
    l_k' & \leq & \hat{l}_k' =  M_0 \Delta \|\bb_k^T A\| e^{(\|A\|\Delta)}\\
    l_{k+i} & \leq & \hat{l}_{k+i} =  M_0 \Delta \|\aa_i^T A\| e^{(\|A\|\Delta)},
               ~~ 1 \leq i \leq k-1 \\
    l_{k+i}' & \leq & \hat{l}_{k+i}' =  M_0 \Delta \|\bb_i^T A\| e^{(\|A\|\Delta)},
               ~~ 1 \leq i \leq k-1 \\
\end{eqnarray*}

\section{ Illustration}

  \subsection{Example 1}
   We first illustrate the schematic algorithm presented in Sec. 3
   with the help of an example taken from \cite{DM98}. Let 
   \[
     \left.
     \begin{array}{lcl}
        \dot{x}  & = & a \\
        \dot{y}  & = & b 
     \end{array}
       \right \} ~~ 
     a,\; b > 0, ~~ (x(0),y(0)) \in X_0 = [0,1] \times [0,1].
   \]
  So, for $t \geq 0$, $x(t) = x(0)+at$ and $y(t) = y(0)+bt$. It is easy
  to check that $S_0^+ = \{1\}\times [0,1] \bigcup\; [0,1]\times\{1\}$.
  Therefore
  \[
 X^+(\tau) = X_0 \bigcup_{0\leq t \leq \tau} \{(1+at,c+bt)\s:~ 0\leq c \leq 1\}
          \bigcup_{0\leq t \leq \tau} \{(c+at,1+bt)\s:~ 0\leq c \leq 1\}.
  \]
  This is illustrated in Fig. 1.

\begin{figure}[h]
\centering
\includegraphics{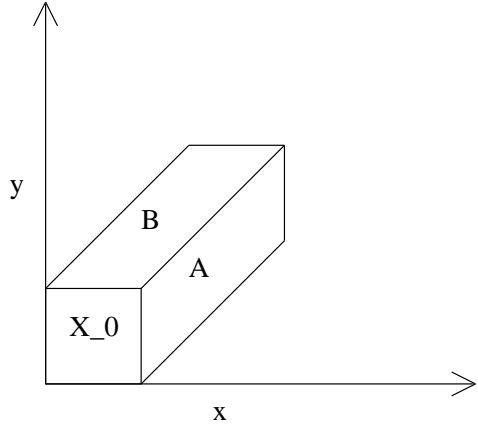}
\caption { Illustration of the Generalized Face Lifting Method: $X^+(\tau) = X_0 \bigcup A \bigcup B$.}
\end{figure}

\subsection{Example 2}

  In this example, we illustrate the over-approximation algorithm.
  Let $\xx(t) = [x_1(t), x_2(t)]^T \in \reals^2$ satisfy 
  \[
    \left.
     \begin{array}{lcl}
        \dot{x}_1  & = & -x_2\\
        \dot{x}_2  & = & x_1
     \end{array}
       \right \} ~~
      (x_1(0),x_2(0)) \in  F_0 = [1,\sqrt{2}] \times \{0\}.
 \]
 The solution is given explicitly by 
\[
 \left[
  \begin{array}{c}
    x_1(t) \\
    x_2(t)
    \end{array}
  \right]
   =
 \left[
  \begin{array}{cc}
    \cos t ~ & ~ -\sin t\\
     \sin t ~ &~~ \cos t
    \end{array}
  \right]
 \left[
  \begin{array}{c}
    x_1(0) \\
    x_2(0)
    \end{array}
  \right]
\]

 With $\Delta = \pi/6$, for the time interval $[0,\pi/6]$,
 the solution set in a parametric form is 
 $T_0 = \{(a\cos t, a\sin t): a \in [1,\sqrt{2}],\; 0 \leq t \leq \pi/6\}$.
 For this example, in the notation of Sec. 4, we have
\[
  \aa_1 = \left[ \begin{array}{c}
                  1\\
		  0
                  \end{array}
           \right], ~~
  \aa_2 = \left[ \begin{array}{c}
                  -1\\
		  0
                  \end{array}
           \right], ~~
  \aa_3 = \left[ \begin{array}{c}
                  0\\
		  1
                  \end{array}
           \right], ~~
  b_1 = \sqrt{2}, ~~ b_2 = -1, ~\textrm{ and  }  b_3 = 0,
\]
 and
\[
  \bb_1 = \left[ \begin{array}{c}
                  \frac{\sqrt{3}}{2}\\
		  0.5
                  \end{array}
           \right], ~
  \bb_2 = \left[ \begin{array}{c}
                  -\frac{\sqrt{3}}{2}\\
		  -0.5
                  \end{array}
           \right], ~
  \bb_3 = \left[ \begin{array}{c}
                  -0.5\\
		  \frac{\sqrt{3}}{2}
                  \end{array}
           \right], ~
  b'_1 = \sqrt{2}, ~ b'_2 = -1, ~\textrm{ and  }  b'_3 = 0.
\]
 The set $T_0$ and the result of the algorithm with $l_i$ and
 $l_i'$ exactly found as in the schematic of Sec. 4 are
 shown in Fig. 2, where the horizontal axis corresponds to
 $x_1$ and the vertical axis corresponds to $x_2$.
\\
 
\begin{figure}[h]
\centering
\centerline{\psfig {file=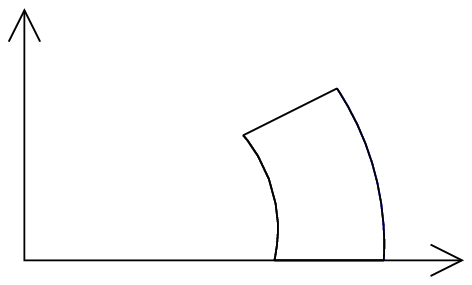} \hspace{0.2in} \psfig {file=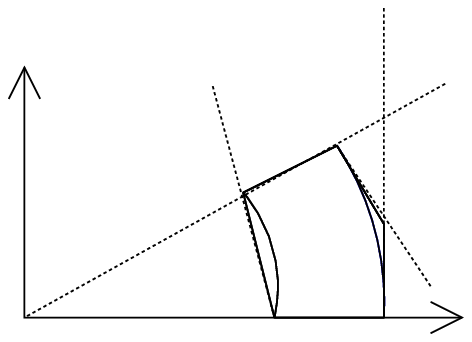}}
\caption {Illustration of over-approximation algorithm of Sec. 4 with Example 2. {\sf Left:} The set $T_0$. {\sf Right:} Result obtained by the schematic algorithm  of Sec. 4, with exact values for $l_i$ and $l_i'$. It may be observed that only $5$ of the $4k = 4 \times 3 = 12$ hyperplanes that the the algorithm computes are required for this example, provided that $l_i$ and $l_i'$ can be found exactly.}

\end{figure}

 Now, recalling the notation of Sec. 4, we may take $\delta = 1$,
 $\delta_0 = \frac{\sqrt{3}}{2}$. It is easy to check that
 $M_0 = \sqrt{2}$ and $\|\aa_i^T A \| = \|\bb_i^T A \| = 1$,
 for $1 \leq i \leq 3$.  Hence
\[
  l_i \leq \hat{l} = \frac{\sqrt{2}\times 1 \times e^{\pi/6}}{\sqrt{3}/2}
          = \frac{2\sqrt{2}\times e^{\pi/6}}{\sqrt{3}} \approx 2.7566424
\]
 and the same upper bound holds for $l_i'$, since
 $\delta_1 = \delta_0$ in this example. For $l_k$ and $l_k'$
 we have the following bounds:
\[
  l_k = l_k' \leq \hat{l}_k = \sqrt{2}\times \frac{\pi}{6} \times e^{\pi/6} \approx 1.249999
\]
 In this example, the same upper bound, given by $\hat{l}_k$, holds
 for $l_{k+i}$ and $l_{k+i}'$. Combining all this, the polyhedron is
 the intersection of the half-spaces given by $\eta_i(\xx) \leq 0$
 and $\eta_i'(\xx) \leq 0$, $1 \leq i \leq 6$,
 where 
 \begin{eqnarray*}
   \eta_1(\xx) = ~ x_1 -  2.7566424\; x_2  - \sqrt{2},  &&
        \eta_1'(\xx) = ~ -0.5122958\; x_1 + 2.8873223\; x_2 - \sqrt{2} , \\
%       \textrm {(corresponding to $\aa_1-l_1\aa_3$ and $\bb_1+l_1'\bb_3$)}\\
   \eta_2(\xx) =~ -x_1 - 2.7566424\;  x_2  + 1.0 ,  &&
      \eta_2'(\xx) =~  -2.2443466\; x_1 + 1.8873223\; x_2 + 1.0,   \\
%      \textrm {(corresponding to $\aa_2-l_2\aa_3$ and $\bb_2+l_2'\bb_3$)}\\
    \eta_3(\xx)  =~ - x_2 , &&  \eta_3'(\xx) =~ - x_1+\sqrt{3} x_2, \\
%                   \textrm {(corresponding to $\aa_3$ and $b_3$)}\\
    \eta_4(\xx)  =~ x_2 - 1.249999, &&  \eta_4'(\xx) =~  0.5\; x_1-\frac{\sqrt{3}}{2} x_2 -1.249999.\\
%                   \textrm {(corresponding to $\aa_3$ and $b_3$)}\\
   \eta_5(\xx) = ~ x_1  - \sqrt{2}-1.249999,  &&
        \eta_5'(\xx) = ~ \frac{\sqrt{3}}{2}\, x_1 + 0.5\, x_2 - \sqrt{2}-1.249999, \\
% \textrm {(corresponding to $\aa_1^T\xx-b_1-l_5$ and $\bb_1^T\xx-b_1'-l_5'$)}\\
%
   \eta_6(\xx) =~ -x_1 + 0.249999, && \eta_6'(\xx) = ~ -\frac{\sqrt{3}}{2}\, x_1 - 0.5\, x_2 + 0.249999. \\
%    \textrm {(corresponding to $\aa_2^T\xx-b_1-l_6$ and $\bb_2^T\xx-b_2'-l_6'$)}\\
 \end{eqnarray*}
 In this example, $\eta_4'(\xx) \leq 0$, $\eta_6(\xx) \leq 0$
 and $\eta_6'(\xx) \leq 0$, are redundant.
\\

\begin{figure}[h]
\centering
\includegraphics{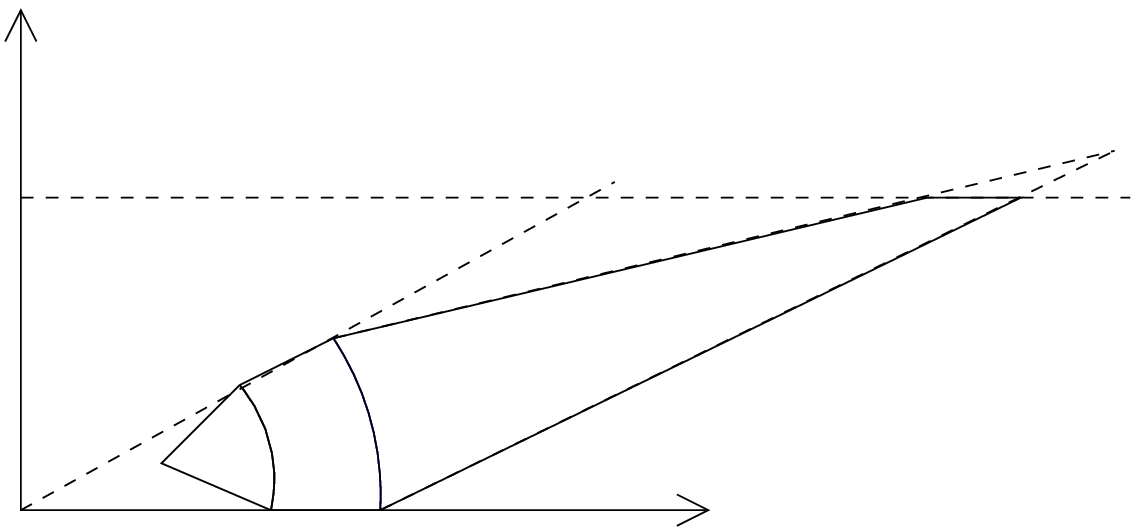}
\caption {\label{FigPoly1}Illustration of polyhedral over-approximation for the soultion set $T_0$ of Example 2: without additional hyperplanes.}
 
\end{figure}

  Fig. \ref{FigPoly1} illustrates the result of the algorithm with
 the first $8$ half-spaces defined by $\eta_i(\xx) \leq 0$
 and $\eta_i' (\xx)  \leq 0$, $1 \leq i \leq 4$, where the dashed lines
 correspond to the lines $\eta_i(\xx) = 0$ and $\eta_i'(\xx) = 0$,
 and the polyhedron included in their intersection is shown in thick lines. 
 The point of intersection of $\eta_1(\xx) = 0$ and
 $\eta_4(\xx) = 0$ is $(4.8600138,\,1.249999)$, of $\eta_4(\xx) =0 $
 and $\eta_1'(\xx) = 0$ is $(4.2845099,\,1.249999)$, and of
 $\eta_2(\xx) = 0$ and $\eta_2' (\xx) = 0$ is $(0.575162,\, 0.154114)$.
 So the vertices in the counter-clockwise order are given by
 $(\sqrt{2},\,0),\;$ $(4.8600138,\,1.249999),\;$ $(4.2845099,\,1.249999),\;$
 $(\sqrt{3}/\sqrt{2},\,1/\sqrt{2}),\;$ $(\sqrt{3}/2,\, 0.5),\;$
 $(0.575162,\, 0.154114)$ and $(1,0)$.
\\

%  Fig. \ref{FigPoly1}, where the dashed lines correspond to the lines
% Additional hyperplanes are
%  given by
% \begin{eqnarray*}
%   \eta_5: ~ x_1  - \sqrt{2}-1.249999,  &&
%        \eta_5': ~ \frac{\sqrt{3}}{2}\, x_1 + 0.5\, x_2 - \sqrt{2}-1.249999, \\
%   \textrm {(corresponding to $\aa_1^T\xx-b_1-l_5$ and $\bb_1^T\xx-b_1'-l_5'$)}\\
%   \eta_6:~ -x_1 + 0.249999, &&
%        \eta_6': ~ -\frac{\sqrt{3}}{2}\, x_1 - 0.5\, x_2 + 0.249999, \\
%    \textrm {(corresponding to $\aa_2^T\xx-b_1-l_6$ and $\bb_2^T\xx-b_2'-l_6'$)}\\
% \end{eqnarray*}
% The point of intersection
%  of $\eta_1(\xx) = 0$ and $\eta_4(\xx) = 0$ is $(4.8600138,\,1.249999)$,
%  of $\eta_4(\xx) =0 $ and $\eta_1'(\xx) = 0$ is $(4.2845099,\,1.249999)$,
%  and of $\eta_2(\xx) = 0$ and $\eta_2' (\xx) = 0$ is $(0.575162,\, 0.154114)$.
%  So the vertices in the counter-clockwise order are given by
%  $(\sqrt{2},\,0),\;$ $(4.8600138,\,1.249999),\;$ $(4.2845099,\,1.249999),\;$
%  $(\sqrt{3}/\sqrt{2},\,1/\sqrt{2}),\;$ $(\sqrt{3}/2,\, 0.5),\;$
%  $(0.575162,\, 0.154114)$ and $(1,0)$.  This is illustrated in
%  Fig. \ref{FigPoly1}, where the dashed lines correspond to the lines
%  $\eta_i(\xx) = 0$, and the polyhedron included in their intersection
%  is shown in thick lines.

\begin{figure}[h]
\centering
\includegraphics{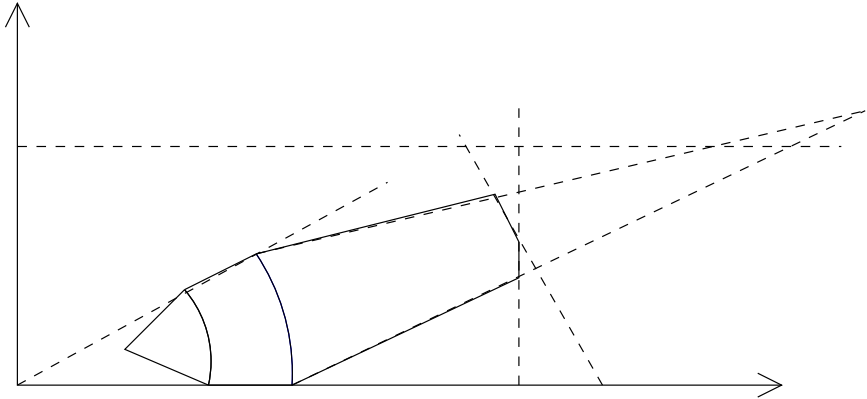}
\caption {\label{FigPoly2} Illustration of polyhedral over-approximation for for the reach set $T_0$ of Example 2: with additional hyperplanes.}
 
\end{figure}

\begin{figure}[h]
\centering
\includegraphics{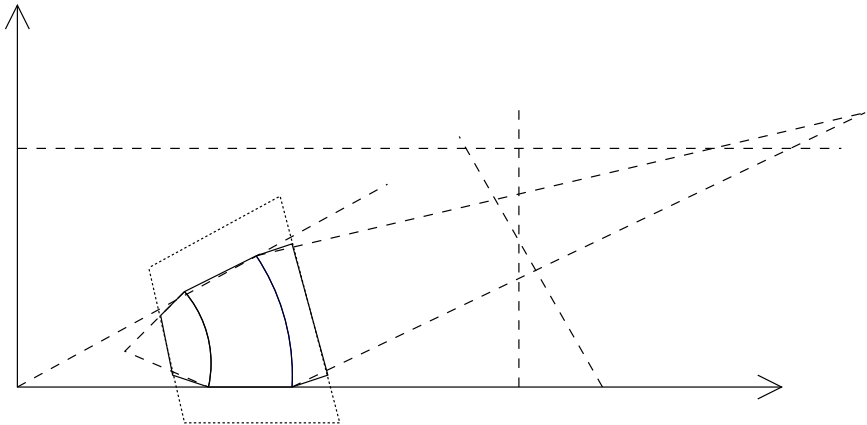}
\caption {\label{FigPoly3} Illustration of polyhedral over-approximation for the solution set $T_0$ of Example 2: intersection of the polyhedron that may be obtained by the method of \cite{ABDM00} with that obtained by the method reported here.}
 
\end{figure}

 Fig. \ref{FigPoly2} shows the result when the remaining two half-spaces
 defined by $\eta_5(\xx) \leq 0$ and $\eta_5'(\xx) \leq 0$ are also used.
 Finally, Fig. \ref{FigPoly3} shows the intersection of the
 polyhedron obtained by the method reported in this paper with
 that which may possibly be obtained by the method of \cite{ABDM00},
 the latter being shown in dotted lines. The polyhedron corresponding
 to the method of \cite{ABDM00} is obtained as follows: the convex
 hull of the sets $[1,\sqrt{2}]\times \{0\}$ and
 $\{(a\cos \frac{\pi}{6}, a\sin \frac{\pi}{6}): a \in [1,\sqrt{2}]\}$
 is the polygon with vertices in the counter-clockwise order
 $(\sqrt{2},0)$, $({\sqrt{3}}/{\sqrt{2}}, 1/{\sqrt{2}})$,
 $({\sqrt{3}}/2, 1/{2})$ and $(1,0)$. Therefore
 the polygon is the interesction of the half-spaces $\zeta_i(\xx) \leq 0$
 and $\zeta_i'(\xx) \leq 0$, $i = 1, 2$, where
 \begin{eqnarray*}
   \zeta_1(\xx)  = - x_2,&& \hspace{-0.2in}
   \zeta_1'(\xx) = - x_1+\sqrt{3} x_2,  ~~\textrm { and}\\
   \zeta_2(\xx)  = \frac{1}{\sqrt{2}} (x_1-\sqrt{2}) -
       \left(\frac{\sqrt{3}}{\sqrt{2}}-\sqrt{2}\right) x_2,&& \hspace{-0.2in}
    \zeta_2'(\xx)  =  -\frac{1}{2} (x_1-1) +
                \left(\frac{\sqrt{3}}{2}-1\right) x_2.
 \end{eqnarray*}
Now if $\hat{\epsilon}$ is an upper bound for the bloating parameter
$\epsilon$, then the half-spaces of the over-approximating polyhedron
corresponding to the method of \cite{ABDM00} are given by
$\hat{\zeta}_i(\xx) \leq 0$ and  $\hat{\zeta}_i'(\xx) \leq 0$,
$i = 1, 2$, where
 \begin{eqnarray*}
   \hat{\zeta}_1(\xx)   & = &  - x_2+\hat{\epsilon}, \\
   \hat{\zeta}_1 '(\xx) & = & - x_1+\sqrt{3} x_2 + \hat{\epsilon}, \\ 
   \hat{\zeta}_2(\xx)  & = & 0.70710678 x_1 + 0.18946869 x_2
                  - 1+\hat{\epsilon}, ~~\textrm { and}\\
    \hat{\zeta}_2'(\xx)  & = &  - 0.70710678 x_1 - 0.1339746 x_2
                  + 0.70710678+\hat{\epsilon}
 \end{eqnarray*}     

\noindent The upper bound for the bloating parameter $\epsilon$
as given in \cite{ABDM00} works out to be
\begin{eqnarray*}
  \epsilon & \leq & M_0 \left(e^{(\|A\|\Delta)} - 1 -
           \|A\|\Delta - \frac{3}{8}\|A\|^2\Delta^2\right)\\
              & = & \sqrt{2} \times \left(e^{(\pi/6)} - 1 - \frac{\pi}{6} - 
                      \frac{3\pi^2}{8\times 36} \right)\\
              & = & \sqrt{2} \times 0.06168464 \approx 0.087235255 < 0.09.
\end{eqnarray*}

 Fig. \ref{FigPoly3} shows the results when an upper bound for the
 bloating parameter is chosen to be $\hat{\epsilon} \approx 0.2$, 
 where the dotted lines correspond to the line $\hat{\zeta}_1 (\xx) = 0$,
 $\hat{\zeta}_1' (\xx) = 0$, $\hat{\zeta}_2 (\xx) = 0$ and
 $\hat{\zeta}_2'(\xx) = 0$. As may be expected, the polyhedron of
 intersection of the two polyhedra -- one polyhedron bounded by the
 dashed lines corresponding to the method described here and another
 bounded by the dotted lines corresponding to the method of
 \cite{ABDM00} -- gives better results of over-approximation.
                                         
%\vspace*{0.2in}

\section{Discussion and Conclusion}

 An important issue of hybrid systems appears to be 
 computability of the reach sets of the continuous variables.
 From a computational point, both the problems of computation
 and efficient representation of the reach sets of the
 continuous varibles are difficult, in general, owing to
 the limitations of quantifier elimination method.
 In this context, approximation of the reach sets by more
 convenient sets, such as polyhedra and subalgebraic sets,
 is discussed in the literature. In this paper, along with
 a method for finding the reach sets, an algorithm for
 over-approximation of the reach sets with polyhedra when
 dynamics of the continuous variables are specified by
 linear differential equations and the inital set is a 
 polyhedron. A practical version of the over-approximation
 algorithm is also discussed in this paper. However, it
 seems that better results of over-approximation may be
 obtained by taking the intersection of the polyhedron
 obtained by the method reported here with that obtained
 by the method given in \cite{ABDM00}. It is hoped that
 the over-approximation method presented here may be
 extended to systems with more general dynamics and
 initial sets.

{
%\section*{Appendix A: Decidable Classes of Subsets of $Q\times\reals^n$}
\section*{Appendix A. Decidable Classes of Subsets of $Q\times\reals^n$}
  
  We introduce here model theoretic concepts for formalizing
  the notion of decidable class $\ccal$ of subsets of
  $Q \times \reals^n$ that we were interested in Sec. \ref{ReachProblems}.
  Since $Q$ is a finite set, for a decidable class $\ccal$ of subsets
  of $Q \times \reals^n$, we may take, for each $q \in Q$, a decidable
  class of subsets of $\reals^n$, $\scal_q$, $q \in Q$, and choose $\ccal$
  to be the product class
  $ \bigcup_{q \in Q} \{ \{q\} \times S_q : S_q \in \scal_q\}$. 
  Therefore, we restrict our attention to the discussion of classes
  of subsets of $\reals^n$. Most parts of this section are taken
  from \cite{Laff99a} (see also \cite{Laff00}, which in turn cites
  the references \cite{Hodges,Marker,vanDalen}).

\begin{mydef}
 A {\em language} is a tuple of three sets, $\lcal = (L_r, L_f, L_c)$,
 where
  \begin{enumerate}
    \item $L_r$ is a set of relations,
    \item $L_f$ is a set of functions, and
    \item $L_c$ are a set of constants.
  \end{enumerate}
\end{mydef}

\begin{mydef}
 A {\em model} of a language $\lcal = (L_r, L_f, L_c)$ consists of a
 nonempty set $\scal$, together with an interpretation of the relations,
 functions and constants.
\end{mydef}

 We denote a model by $(\scal, L_r, L_f, L_c)$, where the
 interpretation is not made explicit. In the following, let
 $\vcal = \{x, y, z, x_0, x_2, \ldots\} $ denote a countable
 set of {\em variables}.

\begin{mydef}
   A {\em term} of a language, $\lcal = (L_r, L_f, L_c)$,
 is inductively defined as follows:
  \begin{enumerate}
   \item each variable $\theta \in \vcal$ is a term,
   \item each constant $c \in L_c$ is a term, and
   \item for an $m$-ary function $g \in L_f$, where $m \geq 1$,
         and $m$ terms, $\theta_1, \theta_2, \ldots, \theta_m$,
         $g(\theta_1, \theta_2, \ldots, \theta_m)$ is a term.
  \end{enumerate}
\end{mydef}
 
\begin{mydef}
   An {\em atomic formula} of a language $\lcal$ is either
  \begin{enumerate}
   \item $\theta_1 = \theta_2$, where $\theta_1$ are $\theta_2$ two terms
         of $\lcal$, or
   \item $\rho(\theta_1, \ldots, \theta_n)$, where $n \geq 1$ and
         $\rho \in L_r$ is an $n$-ary relation.
  \end{enumerate}

\end{mydef}

\begin{mydef}
   A {\em first order formula}, or simply a {\em formula},
 of a language $\lcal$ is recursively defined as one
 of the following:
  \begin{enumerate}
   \item an atomic formula, or

   \item $\lnot \phi$, where $\phi$ is formula and
           $\lnot$ is logical negation, or

   \item $\phi_1 \land \phi_2$, where $\phi_1$ and $\phi_2$ are formulas and
            $\land$ is logical and, or

   \item $\exists x: \phi$ or $\forall x: \phi$, where $\phi$ is a formula,
         $x$ is a variable, and $\exists$ (there exists) and $\forall$
         (for all) are quantifiers; in this case, each occurrence of
         the variable $x$ in the formula $\phi$ is called a {\em bound}
         occurrence. 

  \end{enumerate}

\end{mydef}

\begin{mydef}
 The occurrence of a variable in a formula is {\em free},
 if it is not {\em bound}. A {\em sentence} in a language $\lcal$
 is a formula with no free variable. A {\em theory} of $\lcal$ is
 a subset of sentences.

\end{mydef}

 For a model $\scal$ of the language $\lcal$, we shall be particularly
 interested in the theory defined by the set of all sentences that are
 true in $\scal$. To emphasize this, we refer to this theory as
 {\em the theory of $(\scal, L_r, L_f, L_c)$}.

\begin{mydef}
  Let $\lcal$ be a language and $\scal$ be a model of $\lcal$.
 A set
 $X \subset \scal^n$ is said to be {\em definable} in the language
 $\lcal$, if there is an $n$-ary formula $\phi(x_1,\ldots,x_n)$ such
 that $X$ can be written as
  $X = \{ (x_1,\ldots,x_n) \in S^n: \phi(x_1,\ldots,x_n)\}$.
\end{mydef}

\begin{mydef}
  Let $\lcal$ be a language, $\scal$ be a model of $\lcal$, and
 $\ccal$ be the class of definable sets. Then $\ccal$ is said to
 be {\em decidable}, if the theory of $(\scal, L_r, L_f, L_c)$ is
 decidable, i.e., there is a decision procedure that, given an
 $\lcal$-sentence $\phi$, decides whether $\phi$ belongs to
 the theory of $(\scal, L_r, L_f, L_c)$ or not.
\end{mydef}

\noindent {\bf Examples}

  \begin{enumerate}
   \item The theory $(\reals, \{<\}, \{+,-\}, \{0,1\})$ is the theory
         of linear constraints with integer coefficients, denoted by
         $Lin(\reals)$. The sets defined by these formulas are called
         {\em polyhedral sets}.

   \item The theory $(\reals, \{<\}, \{+,-,\cdot\}, \{0,1\})$ is the theory
         of polynomial constraints with integer coefficients, denoted by
         $OF(\reals)$. The sets defined by these formulas are called
         {\em subalgebraic sets}.
  \end{enumerate}

  The result stated below is due to A. Tarski \cite{Tarski51}:
\begin{myth}
  The first order theory $OF(\reals)$ is decidable.
\end{myth}

\begin{mydef}
  Let $\lcal$ be a language and let $\scal$ be a model of $\lcal$. We 
  say the theory of $(\scal, L_r, L_f, L_c)$ admits {\em quantifier elimination}
  if every first order formula of $(\scal, L_r, L_f, L_c)$ is equivalent
  to a formula of $(\scal, L_r, L_f, L_c)$ without quantifiers.
\end{mydef}

\noindent{ \bf Examples: Decidability and Quantifier Elimination}
 \begin{enumerate}
  \item The theory $OF(\reals)$ consisting of
        $(\reals, \{<\}, \{+,-,\cdot\}, \{0,1\})$
        admits quantifier elimination and is also decidable.
 
  \item Let  $OF_{exp}(\reals)$ be the theory consisting of
         $(\reals, \{<\}, \{+,-,\cdot, exp\}, \{0,1\})$,
        where $exp$, representing the exponential function,
        is a new function symbol. This theory does not admit
        quantifier elimination, and it is not known whether
        this theory is decidable. (See \cite{PappasPhDThesis}).
 
 \end{enumerate}
                                                                                   
}

{
\section*{Appendix B: $Reach_q(X_0,[0,\tau]) = X(\tau) = X^+(\tau)$}

  We assume that $X_0$ is closed and its boundary, denoted by $S_0$,
  consists of a finite union of smooth surfaces. Let
  \[
   X(\tau) = X_0 \bigcup_{0\leq t \leq \tau} \{\phi_q(\xx_0,t)\s:~\xx_0\in S_0\}.
  \]
  Also let
\begin{equation}
  S_0^+ = \{ \xx \in S_0\s: \exists \epsilon = \epsilon(\xx) > 0
    \textrm{ such that } \phi(\xx,t) \in X_0^c, ~ \forall t \in (0,\epsilon) \},
\end{equation}
and $X^+(\tau) = X_0 \bigcup_{0\leq t \leq \tau} \{\phi_q(\xx_0,t)\s:~\xx_0\in S_0^+\}$.

 \begin{myprop}
  $X^+(\tau) = X(\tau) = Reach(X_0,[0,\tau])$.
 \end{myprop}

\noindent {\sf Proof.}
 Note that $X^+(\tau) \subset X(\tau) \subset Reach(X_0,[0,\tau])$.
 So we have to show that  $Reach(X_0,[0,\tau]) \subset X^+(\tau)$. Let
           $\zz \in  Reach(X_0,t)$, $t > 0$.
           If $\zz \in X_0$, then $\zz \in X^+(\tau)$.
           Now suppose $\zz \not \in X_0$. So there is a $\zz_0 \in X_0$
           such that $\zz = \phi(\zz_0,s)$, for some $s$ with $0 < s \leq \tau$.
           If $\zz_0 \in S_0^+$, then $\zz \in X^+(\tau)$.  Otherwise let
 $\tau' = \sup\{t\s:~ t \leq s \textrm{ and } \phi(\zz_0,t) \in  X_0\}$,
 so $0 < \tau' < s \leq \tau$. Now $\zz' = \phi(\zz_0, \tau') \in S_0^+$.
           Therefore, $\zz = \phi(\zz', s-\tau') \in X^+(\tau)$.
\\

  Now suppose that $X_0$ is specified as
  $X_0 = \{\xx \in \reals^n: \ell(\xx) \leq 0\}$, 
  where  $\ell : \reals^n \rightarrow \reals$ is continuously differentiable.
  Further, we assume that if
  $\xx \in {\stackrel{\circ}{X_0}}$ then $\ell(\xx) < 0$
  (where ${\stackrel{\circ}{X_0}}$ denotes the interior of $X_0$,
  i.e., the largest open set contained in $X_0$)
  So, obviously, if $\xx \in S_0$ then $\ell(\xx) = 0$,  and
  if $\xx \in X_0^c$ then $\ell(\xx) > 0$. Let $S_1^+ \subset S_0$
  be defined as 
  \[
      S_1^+ = \{\xx \in S_0\s: ~ \nabla \ell (\xx) \cdot f(q,\xx) \geq 0 \}.
  \]
  We now show that $S_0^+ \subset S_1^+$. To this end, we show
  that $S_0 \backslash S_1^+ \subset S_0 \backslash S_0^+$.
  Let $\xx \in S_0 \backslash S_1^+$, so
  $\nabla \ell (\xx) \cdot f(q,\xx) < 0$. Now, since
  $\ell(\phi_q(\xx,0)) = \ell(\xx) = 0$, and since at $t = 0$,
  $\frac{d\ell(\phi_q(\xx,t))}{dt} = \nabla \ell(\xx)\cdot f(q,\xx) < 0$,
  we have, for a sufficiently small $\epsilon > 0$,
  $\ell(\phi_q(\xx,t)) < 0$, whenever $0 < t < \epsilon$, which
  happens only if $\phi_q(\xx,t) \in {\stackrel{\circ}{X_0}}$,
  $\forall t \in (0,\epsilon)$, implying that $\xx \not \in S_0^+$;
  hence $\xx \in S_0 \backslash S_0^+$. Thus
  $S_0 \backslash S_1^+ \subset S_0 \backslash S_0^+$,
  as required.

}

{

\section*{Appendix C: $Reach_q'\left (X_0,X_q,[0,\infty)\right)$}

  Let $X_q \subset \reals^n$ with compact closure, and let
  $Y_q = \overline{X_q}$. Let $\ff: W \rightarrow \reals^n$ be
  a continuous function defined on an open set $W$ containing
  $Y_q$, satisfying a Lipschitz condition on $W$. Let
  $X_0$ be a closed subset of $Y_q$. We assume
  that for each $\xx \in X_0$, a function
  $\gamma_{\xx} : \reals^+ \rightarrow \reals^n$ exists
  and satisfies the differential equation
  \begin{equation}
          \frac{d\gamma_{\xx}(t)}{dt} = \ff(\gamma_{\xx}),
             ~~ t \geq 0, \label{AppC-Eqn1}
  \end{equation}
  with the initial condition $\gamma_{\xx}(0) = \xx$. Hence the flow
  $\phi(\xx,t)$ associated with equation (\ref{AppC-Eqn1}) is
  defined for all $t \geq 0$ and $\xx \in X_0$. Further, assume
  that the $\omega$-limit set of the flow $\phi(\xx,t)$, $t \geq 0$,
  does not intersect $Y_q$ for any point $\xx \in X_0$. More
  precisely, we assume, for $\xx \in X_0$,
  \begin{equation}
     L_{\omega} (\xx) \bigcap Y_q = \emptyset,  \label{LimitReq}
  \end{equation}
  where 
  \[
   L_{\omega}(\xx) = \bigcap_{t \geq 0}
            \overline{ \phi(\xx, [t,\infty) ) }.
  \]
  (See \cite{Hubbard&West,Hirsch&Smale}.) In what follows, we show
  that if (\ref{LimitReq}) holds for every $\xx \in X_0$, then
  $\exists \tau_{\max} > 0$ (depending on $X_0$), such that
  for every $\xx \in X_0$, there is a $\tau = \tau(\xx)$ with
  $0 < \tau(\xx) \leq \tau_{max}$ and $\phi(\xx, t(\xx)) \not \in Y_q$.
  For this, let  $0 < \tau_1 < \tau_2 < \ldots$ be an 
  increasing sequence such that $\tau_k \rightarrow \infty$,
  as $k \rightarrow \infty$ (as in Sec. 3), and define the sets 
  $A(\tau_k, \xx) = \overline{  \phi(\xx, [\tau_k,\infty) ) }$.
  We first observe that $L_{\omega}(\xx) = \bigcap_k A(\tau_k, \xx)$.

  \begin{myprop}
  \label{AppC-Proposition2}
    Let $\xx \in X_0$.  If (\ref{LimitReq}) holds, then
  $\exists \tau = \tau(\xx) > 0$ such that $\phi(\xx,t) \not \in Y_q$,
  $\forall t \geq \tau$.
  \end{myprop}

  \noindent  {\sf Proof.} We have
  $L_{\omega}(\xx) \bigcap Y_q = $
  $\left [ \bigcap_{k} A(\tau_k,\xx) \right ] \bigcap Y_q$
  $ = \bigcap_{k} \left[ A(\tau_k,\xx) \bigcap Y_q \right ] $. Now let
  $E(\tau_k, \xx) =  A(\tau_k,\xx) \bigcap Y_q $. So, for a fixed
  $\xx \in X_0$, the sets $E(\tau_k,\xx)$, $k = 1, 2, 3, \ldots$,
  is a decreasing sequence of closed subsets of $Y_q$.  Since $Y_q$
  is compact, $\bigcap_{k} E(\tau_k, \xx) = \emptyset$ implies
  $E(\tau_k,\xx) = \emptyset$, for all but finitely many $k$.
  Therefore, for some $K \geq 1$, $\forall k \geq K$,
  $E(\tau_k,\xx) = \emptyset$. Hence,
  $\phi(\xx,[t,\infty)) \bigcap Y_q = \emptyset$, $\forall t \geq \tau_K$. 
  Therefore, with $\tau = \tau_K$, $\phi(\xx, t) \not \in Y_q$,
  $\forall t \geq \tau$.
\\

\noindent We also need the following proposition:

  \begin{myprop}
  \label{AppC-Proposition3}
      Let $\xx \in X_0$ be a point for which there is a $\tau > 0$,
      such that $\phi(\xx, \tau) \not \in Y_q$.  Then there is a
      $\delta = \delta_{\xx} > 0$, such that if $|\xx-\yy| < \delta_{\xx}$
      and $\yy \in X_0$, then $\phi(\yy,\tau) \not \in Y_q$.
  \end{myprop}

  \noindent  {\sf Proof.} 
     Let $|\ff(\xx) - \ff(\yy)| \leq C|\xx-\yy|$. Then
     $|\phi(\xx, \tau) - \phi(\yy, \tau)| \leq |\xx-\yy| e^{\s C\tau}$
  (see, for example, \cite{Hirsch&Smale}), so for a fixed $\tau$,
  $\phi(\cdot,\tau)$ is continuous in the first variable. Let
  $\zz = \phi(\xx, \tau)$. Now since $W \backslash Y_q$ is open, there
  is an $\epsilon > 0$ such that $B(\zz, \epsilon) \subset W \backslash Y_q$.
  By the continuity of $\phi(\cdot,\tau)$ at $\xx$, there is a $\delta > 0$,
  such that
  $|\phi(\xx, \tau) - \phi(\yy,\tau)| = |\zz - \phi(\yy,\tau)| < \epsilon$,
  whenever $|\yy-\xx| < \delta$ and $\yy \in X_0$. Therefore,
  $\phi(\yy,\tau)  \in  B(\zz, \epsilon) \subset W \backslash Y_q$.
\\

\noindent From the previous two propositions, we get the following theorem:

  \begin{myth}
   Let $Y_q$ be compact, and $\ff$ be a continuous function defined
   on an open set $W$ containing $Y_q$ satisfying a Lipschitz condition.
   Further assume condition (\ref{LimitReq}) holds for every
   $\xx \in X_0$. Then $\exists \tau_{\max} > 0$ independent of $\xx$,
   such that for each $\xx \in X_0$, there is a $\tau(\xx)$ with
   $0 < \tau(\xx) \leq \tau_{max}$ and $ \phi(\xx, \tau(\xx)) \not \in Y_q$.
  \end{myth}

  \noindent {\sf Proof.} 
     Let $\xx \in X_0$. By Proposition \ref{AppC-Proposition2}, there is
     a $\tau (\xx) > 0$, such that $\phi(\xx, \tau(\xx)) \not \in Y_q$. By
     Proposition \ref{AppC-Proposition3}, there is a $\delta_{\xx} > 0$,
     such that for any $\yy \in X_0$ with $|\yy-\xx| < \delta_{\xx}$,
     $\phi(\yy, \tau(\xx)) \not \in Y_q$. $\{B(\xx,\delta_{\xx})\s: \xx \in X_0\}$
     is an open cover of $X_0$, containing a finite subcover, say,
     $\{B(\xx_1,\delta_{\xx_1}), \ldots, B(\xx_m,\delta_{\xx_m})\}$.
     Let $\tau_{\max} = \max\{\tau (\xx_1), \ldots, \tau (\xx_m)\}$.
     To check wether this choice of $\tau_{max}$ is as in
     the theorem, let $\yy \in X_0$ be an arbitrary point.
     Now $\yy \in B(\xx_i,\delta_{\xx_i})$, for some
     $i$ with $1 \leq i \leq m$, and by the choice of
     $\delta_{\xx_i}$, $\phi(\yy, \tau(\xx_i)) \not \in Y_q$,
     concluding the proof.

%\\
%  \noindent The significance of this theorem is that the reach set computation
%      algorithm terminates when the solutions are constrained to remain in
%      a compact set.
}

{

\section*{Appendix D: Boundedness Results}

  In this section, we show that if the initial set $F$ is bounded
 then the polyhedron, $P$, enclosing $F_{[0,\Delta]}$, as obtained by
 the algorithm described in Sec. 3 is bounded. We assume that $F$
 is nonempty. Recall that $F$ is given as the set of points $\xx$
 in $\reals^n$ which satisfy the following constraints:
 \begin{equation}
  \left.
   \begin{array}{rcl}
    \aa_i^T \xx - b_i & \leq &  0, ~~i = 1,2, \ldots, k-1, \\
    \aa_k^T \xx - b_k  & = &  0, 
   \end{array}
     \right\}  \label{EqnsForF}
 \end{equation}
 and $P$ is included in the set of points $\xx \in \reals^n$ satisfying
 \begin{equation}
  \left.
   \begin{array}{rcl}
  \aa_i^T \xx-b_i-l_i(\aa_k^T\xx-b_k) & \leq &  0, ~~i = 1,2, \ldots, k-1,\\
   \aa_k^T \xx - b_k  & \geq &   0, ~~\textrm {  and} \\
   \aa_k^T \xx - b_k-l_k & \leq &  0, 
   \end{array}
     \right\}  \label{EqnsForP1}
 \end{equation}
 where $l_i \in \reals$ and $l_k > 0$. Looking at the constraints, one
 may visualize the set $P_1$ satisfying (\ref{EqnsForP1}) as a prism or
 a truncated pyramid, with its bottom given by (\ref{EqnsForF}) and its
 top given by 
 \begin{equation}
  \left.
   \begin{array}{rcl}
    \aa_i^T \xx - b_i - l_i(\aa_k^T\xx-b_k) & \leq &  0, ~~i = 1,2, \ldots, k-1, \\
    \aa_k^T \xx - b_k-l_k  & = &  0.
   \end{array}
     \right\}  \label{EqnsForF_Delta}
 \end{equation}   
  We assume that $F$ is nonempty, and wish to show that if $F$ is bounded
  then so is the set $P_1$, defined as the set of points satisfying
  (\ref{EqnsForP1}).  We start with the following proposition:

 \begin{myprop}
  \label{AppD-Proposition1}
  Assume that $F$ is nonempty and $P_1$ is not bounded.
  Then there is a point $\xx_0 \in F$ and a vector
  $\lambda \in \reals^n$, with $|\lambda| = 1$, such that
  $\xx_0 + t \lambda  \in P_1$, for all $t \geq 0$.
 \end{myprop}

\noindent {\sf Proof.}
 Fix a point $\xx_0 \in F$, and suppose $P_1$ is not bounded.  So
 for each positive integer $k$, there is a point $\xx_k \in P_1$,
 such that $|\xx_k-\xx_0| \geq k$. Let
 $\lambda_k = \frac{(\xx_k-\xx_0)}{|\xx_k-\xx_0|}$. Now, since
 $F \subset P_1$, $\xx_0 \in P_1$ and, by convexity, the entire
 line segment $\xx_0 + t \lambda_k \in P_1$, for $0 \leq t \leq k$.
 In particular, the points on these line segments satisfy (\ref{EqnsForP1}).
 Now, since for each $k$, $|\lambda_k| = 1$, and the closed unit ball
 in $\reals^n$ is compact, there is a convergent subsequence of
 $\lambda_k$'s -- say, $\lambda_{k_i}$, $i = 1, 2, 3, \ldots$ --
 such that $\lambda_{k_i} \rightarrow \lambda \in \reals^n$, as
 $i \rightarrow \infty$; since $|\lambda_{k_i}| = 1$, $|\lambda| = 1$.
 We show that $\xx_0 + t \lambda \in P_1$ for $t \in [0, \infty)$.
 First note that for any positive integer $m$, and for all $i \geq m$,
 $\xx_0+t\lambda_{k_i}$ satisfies the constraints (\ref{EqnsForP1}),
 for $t \in [0,k_m]$. Therefore, $\xx_0+t\lambda$ satisfies
 (\ref{EqnsForP1}), for $t \in [0,k_m]$, and so
 $\xx_0 + t \lambda \in P_1$, for $t \in [0,k_m]$.
 The proposition is concluded by letting $m \rightarrow \infty$. \\

 We now show that such an $\lambda$, as in the previous proposition,
 must be parallel to the hyperplane passing through $F$, the normal
 of which is given by $\aa_k$.

\begin{myprop}
  \label{AppD-Proposition2}
 If $\xx_0 \in F$ and a unit vector $\lambda \in \reals^n$
 are such that $\xx_0 + t \lambda \in P_1$, for all
 $t \in [0,\infty)$, then $\aa_k^T\lambda = 0$.
\end{myprop}

\noindent {\sf Proof.}
 For a contradiction assume that $h = \aa_k^T\lambda > 0$. Now
 since $\xx+t\lambda$ satisfies (\ref{EqnsForP1}), we must have
 \[
  \aa_k^T (\xx_0+t\lambda) - b_k-l_k  \leq   0,
 \]
 which holds only if $t \leq \frac{(b_k+l_k-\aa_k^T\xx_0)}{h}$,
 contrary to the hypothesis that, for all $t \in [0,\infty)$,
 $\xx_0 + t \lambda \in P_1$. Similarly, if $h = \aa_k^T\lambda < 0$,
 then the constraint
\[
  \aa_k^T (\xx_0+t\lambda) - b_k  \geq   0,
\]  
 does not hold for $t > 0$. Therefore we must have
 $\aa_k^T\lambda = 0$. \\

 We now show that if $F$ is bounded, then for any unit vector $\lambda$
 orthogonal to $\aa_k$, there is an $i$ with $1 \leq i \leq k-1$, such that
 $h_i = \aa_i^T\lambda > 0$.

\begin{myprop}
  \label{AppD-Proposition3}
  If $F$ is nonempty and bounded, then for any unit vector
  $\lambda \in \reals^n$, for which $\aa_k^T \lambda = 0$, there
  is an $i$, with $1 \leq i \leq k-1$, such that $\aa_i^T \lambda > 0$.
\end{myprop}

\noindent {\sf Proof.}
  Let $\xx_0 \in F$, and suppose that there is an $\lambda \in \reals^n$,
 with $|\lambda| = 1$, $\aa_k^T \lambda = 0$ and $\aa_i^T \lambda \leq 0$,
 for $1 \leq i \leq k-1$. Then for any $t > 0$, the vector
 $\xx_0 + t \lambda$ satisfies the constraints (\ref{EqnsForF}), and
 so $\xx_0 + t \lambda \in F$, contrary to the assumption that $F$
 is bounded. \\

  We now combine all the previous propositions to get the following
 result:

 \begin{myth}
  If $F$ is nonempty and bounded, then $P_1$ is bounded.
 \end{myth}
\noindent {\sf Proof.}
  If $P_1$ is not bounded, then by Proposition \ref{AppD-Proposition1},
  there is an $\xx_0 \in F$ and a unit vector $\lambda \in \reals^n$
  such that $\xx_0 + t \lambda \in P_1$, for any $t \geq 0$.
  By Proposition \ref{AppD-Proposition2}, $\aa_k^T \lambda = 0$. Now,
  by the previous proposition, since $F$ is bounded, there is an $i$,
  with $1 \leq i \leq k-1$, such that $h_i = \aa_i^T\lambda > 0$. But
  then the constraint
  \[
  \aa_i^T (\xx_0+t\lambda)-b_i-l_i(\aa_k^T(\xx_0+t\lambda)-b_k) \leq  0,
  \]
  cannot hold for any $t$ with
  \[
    t > -\frac{\aa_i^T\xx_0 - b_i}{h_i},
  \]
  contrary to the assumption that $\xx_0 + t \lambda \in P_1$,
  for any $t \geq 0$. Hence $P_1$ is bounded.

}

\end{document}